\documentclass[a4paper,11pt]{article}
\pdfoutput=1 

\usepackage{jheppub} 
\makeatletter
\DeclareRobustCommand*{\bfseries}{%
  \not@math@alphabet\bfseries\mathbf
  \fontseries\bfdefault\selectfont
  \boldmath
}
\makeatother

\usepackage{calc}
\usepackage{rotating}
\usepackage[english]{babel}
\usepackage{graphicx}
\usepackage{subfig}
\usepackage{float}
\usepackage{amsmath}
\usepackage{amssymb}
\usepackage{amsthm}
\usepackage{latexsym}
\usepackage{dcolumn}

\newcommand{\newc}{\newcommand*}

\long\def\begincomment#1\endcomment{%
        \begingroup\sf\baselineskip12pt#1\endgroup}

\newc{\etal}{\textrm{et al.}} 
\newc{\eg}{\textrm{e.g.}} 
\newc{\ie}{\textrm{i.e.}}
\newc{\etc}{\textrm{etc}}
\newc\vs{\textrm{vs.}}
\newc{\cl}{\rm {CL}}

\newc{\ev}{\ensuremath{\,\mathrm{eV}}}
\newc{\kev}{\ensuremath{\,\mathrm{keV}}}
\newc{\mev}{\ensuremath{\,\mathrm{MeV}}}
\newc{\gev}{\ensuremath{\,\mathrm{GeV}}}
\newc{\tev}{\ensuremath{\,\mathrm{TeV}}}
\newc{\MeV}{\mev} 
\newc{\TeV}{\tev}
\newc{\invpb}{\ensuremath{/\text{pb}}}
\newc{\invfb}{\ensuremath{/\text{fb}}}
\newc\nb{\ensuremath{\,\mathrm{nb}}} \newc\pb{\ensuremath{\,\mathrm{pb}}} \newc\fb{\ensuremath{\,\mathrm{fb}}}
\newc\pc{\ensuremath{\,\mathrm{pc}}}
\newc\kpc{\ensuremath{\,\mathrm{kpc}}}
\newc\mpc{\ensuremath{\,\mathrm{Mpc}}}
\newc\ps{\ensuremath{\,\mathrm{ps}}} 
\newc\cmeter{\ensuremath{\,\mathrm{cm}}} 
\newc\meter{\ensuremath{\,\mathrm{m}}} 
\newc\kmeter{\ensuremath{\,\mathrm{km}}}
\newc\second{\ensuremath{\,\mathrm{s}}}
\newc\msecond{\ensuremath{\,\mathrm{ms}}}
\newc\nsecond{\ensuremath{\,\mathrm{ns}}}
\newc\psecond{\ensuremath{\,\mathrm{ps}}}

\newc{\chisqmin}{\ensuremath{\chi^2_{\mathrm{min}}}}
\newc{\Delchisq}{\ensuremath{\Delta\chi^2}}
\newc{\chisq}{\ensuremath{\chi^2}}
\newc{\like}{\ensuremath{\mathcal{L}}}

\newc\lsim{\ensuremath{\mathrel{\rlap{\lower4pt\hbox{\hskip1pt$\sim$}}\raise1pt\hbox{$<$}}}}
\newc\gsim{\ensuremath{\mathrel{\rlap{\lower4pt\hbox{\hskip1pt$\sim$}}\raise1pt\hbox{$>$}}}}
\newc{\VEV}[1]{\ensuremath{\langle #1 \rangle}}
\newc{\dl}{\ensuremath{\stackrel{\leftarrow}{D}}}
\newc{\dr}{\ensuremath{\stackrel{\rightarrow}{D}}}

\newc{\bcenter}{\begin{center}}    \newc{\ecenter}{\end{center}}
\newc{\bfl}{\begin{flushleft}}    \newc{\efl}{\end{flushleft}}
\newc{\bfr}{\begin{flushright}}    \newc{\efr}{\end{flushright}}

\newc{\bi}{\begin{itemize}}
\newc{\ei}{\end{itemize}}
\newc{\bed}{\begin{description}}
\newc{\eed}{\end{description}}
\newc{\ben}{\begin{enumerate}}
\newc{\een}{\end{enumerate}}

\newc{\be}{\begin{equation}}
\newc{\ee}{\end{equation}}
\newc{\bea}{\begin{eqnarray}}
\newc{\eea}{\end{eqnarray}}
\newc{\ra}{\rightarrow}

\newc{\alphas}{\ensuremath{\alpha_s}}
\newc{\alphatwo}{\ensuremath{\alpha_2}}
\newc{\alphaone}{\ensuremath{\alpha_1}}
\newc{\alphai}[1]{\ensuremath{\alpha_{#1}}}
\newc{\alphaem}{\ensuremath{\alpha_{\mathrm{em}}}}
\newc{\alphaeff}{\ensuremath{\alpha_{\mathrm{eff}}}}
\newc{\sineff}{\ensuremath{\sin \theta_{\mathrm{eff}}}}
\newc{\sinsqeff}{\ensuremath{\sin^2 \theta_{\mathrm{eff}}}}
\newc{\dalphahad}{\ensuremath{\Delta \alpha_{\mathrm{had}}}}
\newc{\yt}{\ensuremath{h_t}} \newc{\yb}{\ensuremath{h_b}} \newc{\ytau}{\ensuremath{h_{\tau}}}
\newc\mz{\ensuremath{M_Z}} 
\newc\mw{\ensuremath{m_W}}
\newc\mZ{\mz}        \newc\mW{\mw}
\newc\mhsm{\ensuremath{ m_{H_{\mathrm{SM}}}}}
\newc{\mtop}{\ensuremath{ m_t}}               \newc{\mtpole}{\ensuremath{ M_t}}
\newc{\mbottom}{\ensuremath{ m_b}} 
\newc{\mtau}{\ensuremath{ m_{\tau}}}
\newc{\mt}{\mtpole}
\newc{\mb}{\mbottom} 
\newc{\rtwogg}{\ensuremath{R_{h_2}(\gamma\gamma)}}
\newc{\rtwozz}{\ensuremath{R_{h_2}(ZZ)}}
\newc{\ronegg}{\ensuremath{R_{h_1}(\gamma\gamma)}}
\newc{\ronezz}{\ensuremath{R_{h_1}(ZZ)}}
\newc{\rsiggg}{\ensuremath{R_{h_\textrm{sig}}(\gamma\gamma)}}
\newc{\rsigzz}{\ensuremath{R_{h_\textrm{sig}}(ZZ)}}
\newc{\llbar}{\ensuremath{\ell\bar{\ell}}}
\newc{\tauptaum}{\ensuremath{ \tau^+\tau^-}}
\newc{\qqbar}{\ensuremath{ q\bar{q}}} \newc{\ppbar}{\ensuremath{ p\bar{p}}}
\newc{\bbbar}{\ensuremath{ b\bar{b}}} \newc{\ttbar}{\ensuremath{ t\bar{t}}}
\newc{\ffbar}{\ensuremath{ f\bar{f}}} \newc{\tautaubar}{\ensuremath{ \tau\bar{\tau}}}
\newc{\mchi}{\ensuremath{m_\neutone}}
\newc{\squark}{\ensuremath{\tilde{q}}}
\newc{\slepton}{\ensuremath{\tilde{l}}}
\newc{\gluino}{\ensuremath{\tilde{g}}} 
\newc{\mgluino}{\ensuremath{{m_{\gluino}}}}

\newc{\sthw}{\ensuremath{ \sin\theta_W}}              \newc{\cthw}{\ensuremath{\cos\theta_W}}
\newc{\tanthw}{\ensuremath{ \tan\theta_W}}              \newc{\cotthw}{\ensuremath{\cot\theta_W}}
\newc{\ssqthw}{\ensuremath{\sin^2 \theta_W}}
\newc{\msbar}{\ensuremath{\overline{MS}}} \newc{\drbar}{\ensuremath{\overline{DR}}}
\newc{\mtmtsmmsbar}{\ensuremath{ m_t(m_t)^{\msbar}_{{\mathrm{SM}}}}}
\newc{\mtmtsmdrbar}{\ensuremath{ m_t(m_t)^{\drbar}_{{\mathrm{SM}}}}}
\newc{\mtmtmssmdrbar}{\ensuremath{ m_t(m_t)^{\drbar}_{{\mathrm{SUSY}}}}}
\newc{\mbmbmsbar}{\ensuremath{ m_b(m_b)^{\msbar} }}
\newc{\mbmbsmmsbar}{\ensuremath{ m_b(m_b)^{\msbar}_{{\mathrm{SM}}}}}
\newc{\mbmzsmmsbar}{\ensuremath{ m_b(\mz)^{\msbar}_{{\mathrm{SM}}}}}
\newc{\mbmzsmdrbar}{\ensuremath{ m_b(\mz)^{\drbar}_{{\mathrm{SM}}}}}
\newc{\mbmzmssmdrbar}{\ensuremath{ m_b(\mz)^{\drbar}_{{\mathrm{SUSY}}}}}
\newc{\mtaumzsmmsbar}{\ensuremath{ m_{\tau}(\mz)^{\msbar}_{{\mathrm{SM}}}}}
\newc{\mtaumzsmdrbar}{\ensuremath{ m_{\tau}(\mz)^{\drbar}_{{\mathrm{SM}}}}}
\newc{\mtaumzmssmdrbar}{\ensuremath{ m_{\tau}(\mz)^{\drbar}_{{\mathrm{SUSY}}}}}
\newc{\alphasmzms}{\ensuremath{\alpha_s(M_Z)^{\overline{MS}}}}
\newc{\alphaimzms}[1]{\ensuremath{\alpha_{#1}(M_Z)^{\overline{MS}}}}
\newc{\alphaemmz}{\ensuremath{\alpha_{\mathrm{em}}(M_Z)^{\overline{MS}}}}

\newc{\mzero}{\ensuremath{{m_0}}}
\newc{\mhalf}{\ensuremath{ m_{1/2}}}
\newc{\tanb}{\ensuremath{\tan\beta}}
\newc{\azero}{\ensuremath{ A_0}}
\newc{\sgnmu}{\ensuremath{\textrm{sgn}\,\mu}}
\newc{\atau}{\ensuremath{{A_{\tau}}}}
\newc{\mueff}{\ensuremath{\mu_{\rm{eff}}}}
\newc{\lam}{\ensuremath{{\lambda}}}
\newc{\kap}{\ensuremath{{\kappa}}}
\newc{\alam}{\ensuremath{{A_{\lambda}}}}
\newc{\akap}{\ensuremath{{A_{\kappa}}}}
\newc{\hs}{\ensuremath{ H_s}}      
\newc{\mhs}{\ensuremath{ m_{H_s}}} 
\newc{\mgut}{\ensuremath{ M_{\rm GUT}}}
\newc{\mplanck}{\ensuremath{ M_{\rm P}}}      \newc{\mpl}{\ensuremath{ M_{\rm Pl}}}
\newc{\msusy}{\ensuremath{ M_{\rm SUSY}}}      \newc{\ms}{\ensuremath{ M_{\rm S}}}
 \newc{\hu}{\ensuremath{ H_u}}       \newc{\hd}{\ensuremath{ H_d}}
 \newc{\mhu}{\ensuremath{ m_{H_u}}}       \newc{\mhd}{\ensuremath{ m_{H_d}}}
 \newc{\mhuew}{\ensuremath{ m^{\ast}_{H_u}}}       \newc{\mhdew}{\ensuremath{ m^{\ast}_{H_d}}}
 \newc{\mhuewsq}{\ensuremath{ m^{\ast\, 2}_{H_u}}}       \newc{\mhdewsq}{\ensuremath{ m^{\ast\, 2}_{H_d}}}
 \newc{\mhl}{\ensuremath{m_\hl}} 
 \newc{\mhone}{\ensuremath{m_{h_1}}} 
 \newc{\mhtwo}{\ensuremath{m_{h_2}}} 
 \newc{\mglu}{\ensuremath{m_{\tilde g}}} 
 \newc{\mul}{\ensuremath{m_{\tilde{u}_L}}} 
 \newc{\mstopone}{\ensuremath{m_{\tilde{t}_1}}} 
 \newc{\ma}{\ensuremath{m_A}} 
 \newc{\maone}{\ensuremath{m_{a_1}}} 
 \newc{\matwo}{\ensuremath{m_{a_2}}}
 \newc{\hone}{\ensuremath{h_1}}
 \newc{\htwo}{\ensuremath{h_2}}
 \newc{\aone}{\ensuremath{a_1}}
 \newc{\atwo}{\ensuremath{a_2}}

\newc{\sigsip}{\ensuremath{\sigma^{\rm SI}_{p}}}	\newc{\sigsin}{\ensuremath{\sigma^{\rm SI}_{n}}}
\newc{\sigsdp}{\ensuremath{\sigma^{\rm SD}_{p}}}	\newc{\sigsdn}{\ensuremath{\sigma^{\rm SD}_{n}}}
\newc{\sigsi}{\ensuremath{\sigma^{\rm SI}}}	\newc{\sigsd}{\ensuremath{\sigma^{\rm SD}}}
\newc{\abund}{\ensuremath{ \Omega h^2}}
\newc{\omegadm}{\ensuremath{ \Omega_{{\rm DM}}}}     \newc{\abunddm}{\ensuremath{ \Omega_{{\rm DM}} h^2}} 
\newc{\omegam}{\ensuremath{ \Omega_{{\rm m}}}}       \newc{\abundm}{\ensuremath{ \Omega_{{\rm m}} h^2}}
\newc{\omegab}{\ensuremath{ \Omega_{{\rm b}}}}	\newc{\abundb}{\ensuremath{ \Omega_{{\rm b}} h^2}}
\newc{\omegatot}{\ensuremath{ \Omega_{{\rm TOT}}}}
\newc{\omegacdm}{\ensuremath{ \Omega_{{\rm CDM}}}}   \newc{\abundcdm}{\ensuremath{ \Omega_{{\rm CDM}} h^2}}
\newc{\omegalambda}{\ensuremath{ \Omega_{\Lambda}}} \newc{\abundlambda}{\ensuremath{ \Omega_{\Lambda} h^2}}
\newc{\omegarad}{\ensuremath{ \Omega_{{\rm rad}}}}  \newc{\abundrad}{\ensuremath{ \Omega_{{\rm rad}} h^2}}
\newc{\rhocrit}{\ensuremath{ \rho_{\rm crit}}}
\newc{\rhochi}{\ensuremath{ \rho_{\chi}}}
\newc{\abunchi}{\ensuremath{\Omega_\chi h^2}}
\newc{\abundlsp}{\ensuremath{\Omega_{\rm LSP}h^2}}
\newcommand*{\abundchi}{\ensuremath{\Omega_\chi h^2}}

\newc{\amu}{\ensuremath{ a_{\mu}}}        \newc{\amususy}{\ensuremath{ a_{\mu}^{\mathrm{SUSY}}}}
\newc{\amuexpt}{\ensuremath{ a_{\mu}^{\mathrm{expt}}}}        \newc{\amusm}{\ensuremath{ a_{\mu}^{\mathrm{SM}}}}
\newc\deltaamu{\ensuremath{\Delta a_{\mu}}} \newc{\deltaamususy}{\ensuremath{\delta a_{\mu}^{\mathrm{SUSY}}}}
\newc\gmtwo{\ensuremath{ (g-2)_{\mu}}} 
\newc{\deltagmtwomususy}{\ensuremath{\delta\left(g-2\right)_{\mu}^{\mathrm{SUSY}}}}
\newc{\deltagmtwomu}{\ensuremath{\delta\left(g-2\right)_{\mu}}}
\newc\BR{\ensuremath{\textrm{BR}}}
\newc\bsgamma{\ensuremath{ b\rightarrow s \gamma }}
\newc\bxsgamma{\ensuremath{\overline{B}\rightarrow X_{s}\gamma}}
\newc\brbsgamma{\ensuremath{\BR\left(\bsgamma\right)}}
\newc\brbxsgamma{\ensuremath{\BR\left(\bxsgamma\right)}}
\newc\bsmumu{\ensuremath{B_s\to\mu^+\mu^-}}
\newc\brbsmumu{\ensuremath{\BR\left(B_s\to\mu^+\mu^-\right)}}
\newc\bdmmumu{\ensuremath{\overline{B}_d\to\mu^+\mu^-}}
\newc\bbbarmix{\ensuremath{\overline{B}_s\mbox{-}B_s}}      
\newc\delmbs{\ensuremath{\Delta M_{B_s}}}
\newc{\butaunu}{\ensuremath{B_u \rightarrow \tau \nu}}
\newc{\brbutaunu}{\ensuremath{\BR\left(B_u \rightarrow \tau \nu\right)}}

\newcommand*{\reftable}[1]{Table~\ref{#1}}

     \newcommand*{\refsec}[1]{Sec.~\ref{#1}}

\newcommand*{\neutone}{\ensuremath{\chi}}

\newcommand*{\seven}{\ensuremath{\sqrt{s}=7\tev}}
\newcommand*{\eight}{\ensuremath{\sqrt{s}=8\tev}}
\newcommand*{\alphaT}{\ensuremath{\alpha_T}}

\newcommand*{\alphaTelefb}{\ensuremath{\cms\ \alphaT\ 11.7\invfb} }
\newcommand*{\razor}{\textrm{razor}}
\newcommand*{\razorfourfb}{\ensuremath{\cms\ \razor\ 4.4\invfb} }

\newcommand*{\softsusy}{SOFTSUSY}

\newcommand*{\micromegas}{MicrOMEGAs}

\newcommand*{\cms}{\text{CMS}}

\newcommand*{\superiso}{\text{SuperIso}}

\let\oldcite\cite
\renewcommand*{\cite}{~\oldcite}

\newcommand*{\hl}{\ensuremath{h}}

\newcommand*{\ha}{\ensuremath{A}}
\newcommand*{\mha}{\ensuremath{m_\ha}}

\restylefloat{figure}

\title{Two ultimate tests of constrained supersymmetry}

\author{Kamila Kowalska,} 
\author[1]{Leszek Roszkowski\note{On leave of absence from the University of Sheffield, U.K.}}
\author{and Enrico Maria Sessolo}

\affiliation{National Centre for Nuclear Research,
  Ho{\. z}a 69, 00-681 Warsaw, Poland} 

\emailAdd{Kamila.Kowalska@fuw.edu.pl}
\emailAdd{L.Roszkowski@sheffield.ac.uk}
\emailAdd{Enrico-Maria.Sessolo@fuw.edu.pl}

\abstract{We examine the prospects of using two alternative and
  complementary ways to explore the regions that are
  favored by global constraints in two simple unified
  supersymmetric models: the CMSSM and the NUHM. First, we
  consider \brbsmumu, which has recently been for the first time
  measured by LHCb. In the CMSSM we show that ultimate, but realistic,
  improvement in the determination of the observable to about 5-10\%
  around the Standard Model value would strongly disfavor the
  $A$-funnel region, while not affecting much the other favored
  regions. Second, we show that all the favored regions of the CMSSM
  will be, for the most part, sensitive to direct dark matter searches
  in future one-tonne detectors. A signal at low WIMP mass
  ($\lsim450\gev$) and low spin-independent cross section would then
  strongly favor the stau coannihilation region while a signal at higher WIMP
  mass ($\sim800\gev$ to $\sim1.2\tev$) would clearly point to the
  region where the neutralino is higgsino-like
  with mass $\sim1\tev$.  A nearly complete experimental testing of the CMSSM
  over multi-TeV ranges of superpartner masses, 
  far beyond the reach of direct SUSY searches at the LHC, can
  therefore be achievable. In the NUHM, in contrast, similar favored
  regions exist but a sample study
  reveals that even a precise determination of \brbsmumu\ would have a
  much less constraining power on the model, including the $A$-funnel
  region. On the other hand, this could allow one to, by detecting in
  one-tonne detectors a signal for $500\gev\lsim \mchi\lsim 800\gev$,
  strongly disfavor the CMSSM.}

\begin{document}
\maketitle
\flushbottom

\section{Introduction}\label{intro:sec}

In November 2012 the LHC reached the end of its current data
collecting phase with the proton-proton beam at \eight. A huge
amount of data was collected, allowing the CMS and ATLAS
collaborations to reach an integrated luminosity of around 23\invfb\
each, and LHCb of around 2.2\invfb. The performance of the detectors
at the LHC and the effort of the experimental collaborations have been
quite spectacular.  The past year brought some experimental results
whose crucial importance cannot be questioned, even though they still
require further investigation and confirmation with larger amounts of data.

Most notably, on July 4, 2012, both the CMS and ATLAS collaborations
announced a $5\sigma$ discovery of a particle consistent with the
Higgs boson predicted by the Standard Model (SM) based on the analysis
of 4.9\invfb\ of $pp$ collisions at
\seven\cite{CMS:2012gu,ATLAS:2012gk}. Both collaborations have
recently updated their results, combining data from the \seven\ and
\eight\ runs.  The CMS value of the Higgs-like boson mass,
$125.8\pm0.6\gev$\cite{cmshiggs}, is based on the analysis
of the data corresponding to integrated luminosities of 5.1/\fb\ at
$\sqrt{s}=7\tev$ and up to 12.2/\fb\ at $\sqrt{s}=8\tev$ in the
$\gamma\gamma$, $ZZ$, $WW$, $\tau\tau$ and $bb$ decay channels. The
ATLAS analysis combined approximately 4.8/\fb\ of data at
$\sqrt{s}=7\tev$ with 5.8/\fb\ of data at $\sqrt{s}=8\tev$ in the same
five channels, obtaining $125.2\pm0.7\gev$\cite{ATLAS-CONF-2012-170}.

On November 13, 2012 the LHCb Collaboration reported the first evidence 
of an excess in the rare decay \bsmumu\cite{Aaij:2012ct}.
The measured value of the branching ratio,
$\brbsmumu=\left(3.2^{+1.5}_{-1.2}\right)\times10^{-9}$, is consistent
with the value predicted by the SM. This decay has been long
considered as one of the best probes for new physics, and in
particular for low-scale supersymmetry (SUSY), since SUSY
contributions can be largely enhanced by the sixth power of $\tanb$,
the ratio of the vacuum expectation values of the two Higgs doublets
(see,
\eg,\cite{Huang:1998vb,Hamzaoui:1998nu,Choudhury:1998ze,Babu:1999hn,Huang:2000sm,Ellis:2005sc}
for some early studies).

The agreement of the recent measurement with the SM makes it
potentially strongly constraining for the allowed parameter space of
SUSY models. On the other side, the result still suffers from
substantial experimental uncertainties -- its current $2\sigma$ upper
bound is actually weaker than the previous 95\% confidence level (CL)
exclusion limit $\brbsmumu<4.5\times10^{-9}$ obtained earlier by the
same collaboration\cite{Aaij:2012ac}.

Finally, on the front of direct SUSY searches, the \seven\ and \eight\
runs have significantly improved the limits on the masses of colored
superpartners, allowing this way both CMS and ATLAS to exclude
increasingly larger ranges of parameters of low-energy SUSY models.
Currently the most constraining 95\% \cl\ exclusion limits on
the parameter space of the Constrained Minimal Supersymmetric Standard
Model (CMSSM) comes from the ATLAS search for squarks and gluinos with
jets and missing transverse energy in the final states, with
5.8\invfb\ of data at \eight\cite{ATLAS-CONF-2012-109}.  A similar
analysis by CMS based on 11.7\invfb\ of data and using the kinematical
variable \alphaT\ as a discriminator is slightly less
constraining\cite{newalphat}.  On the other hand, as we will show in
this paper, the CMS razor analysis at \seven\ with 4.4\invfb\cite{Chatrchyan:2012uea} can be combined with the
most recent \alphaT\ at \eight\cite{newalphat} to produce a lower bound on the mass
parameters of the CMSSM that, in the regions favored by the global
constraints, is comparable to the current one from ATLAS.

In our recent global Bayesian analysis of the
CMSSM\cite{Fowlie:2012im} (as well as in several other recent global,
Bayesian or \chisq-based,
analyses\cite{Kadastik:2011aa,Balazs:2012qc,Bechtle:2012zk,Akula:2012kk,
  Beskidt:2012sk,Buchmueller:2012hv,Strege:2012bt,Cabrera:2012vu,Citron:2012fg}),
\footnote{A note of caution is in order regarding a quantitative
  comparison of different analyses. First, Bayesian posterior high
  probability credible regions and \chisq\ confidence regions need not
  agree as they are based on two different concepts of
  probability. Secondly, even within the same statistical framework,
  numerical results often strongly depend on the values of input
  parameters used. For instance, the mass of the lightest Higgs boson
  in SUSY very sensitively depends on the exact value of the top quark
  pole mass, which is different in, \eg,\cite{Fowlie:2012im}
  and\cite{Strege:2012bt}. } it was shown or reiterated that, when
combining through the likelihood function the \brbsmumu\ bound from
Ref.\cite{Aaij:2012ac}, the Higgs mass, limits from direct SUSY
searches, the relic density of dark matter (DM), an excess in the
anomalous magnetic moment of the muon \gmtwo, and other relevant
constraints, four clearly identifiable regions of the
(\mzero, \mhalf) plane, with \mzero\ and \mhalf\ denoting the scalar
and gaugino soft masses, respectively, remain favored by high
posterior probability for both signs of the Higgs/higgsino mass
parameter $\mu$:

(a) at small $\mzero\lsim400\gev$ and $600\gev\lsim \mhalf\lsim
1000\gev$, where the correct relic abundance is obtained via efficient
neutralino-stau coannihilation\cite{Ellis:1998kh} (stau-coannihilation
(SC) region hereafter).  In this region the lightest bino-like neutralino
as the lightest SUSY particle (LSP)  is fairly light, $\mchi\lsim450\gev$, and so is the lightest stop,
$\mstopone\sim1\tev$, hence the correct Higgs mass is achieved due to
maximal stop mixing, $A_t^2/m_{\tilde{t}}^2\sim 6$;

(b) at $1\tev\lsim\mhalf\lsim2\tev$, where the cross-section for neutralino
annihilation is enhanced by the $s$-channel resonance of the
pseudoscalar $A$ Higgs boson\cite{Drees:1992am} ($A$-funnel (AF)
region), with bino-like LSP in the mass range $350\gev\lsim\mchi\lsim700\gev$; 

(c) for $\mzero\gsim3\tev$, $\mzero>\mhalf$, in a strip of the (\mzero, \mhalf) plane along the border
of the non-electroweak symmetry-breaking region, where the neutralino
remains bino dominated but 
contains a non-negligible higgsino component\cite{Chan:1997bi,Feng:1999zg} (Focus
Point/Hyperbolic Branch (FP/HB) region). In the FP/HB region we found\cite{Fowlie:2012im}
a significantly lower posterior probability primarily because it was
difficult to obtain the correct mass of the Higgs boson. Also, this
region is in considerable tension with current 90\% \cl\ upper
bound from XENON100\cite{Aprile:2012nq} on the spin-independent cross
section \sigsip\ of dark matter (DM) scattering off xenon nuclei.

(d) in the multi-TeV regime ($\mzero\gsim4\tev$, $\mhalf\gsim2\tev$) there
is a large region where the neutralino LSP is almost purely
higgsino-like\cite{Akula:2012kk}. Its mass is almost constant,
$\mchi\approx\mu\simeq1\tev$ (1TH region hereafter) so that the relic
density constraint is easily satisfied, since for such a heavy
higgsino LSP coannihilation is no longer effective.

A similar pattern holds also in the Non-Universal Higgs Model (NUHM),
although at somewhat different locations. Specifically, the 1TH region
can be found already at much lower mass scales, $\mzero\lsim4\tev$ and
$\mhalf\lsim2\tev$\cite{Roszkowski:2009sm}. (For an updated analysis
including LHC data, see\cite{Strege:2012bt}.)

Clearly, given such large mass scales most of the favored regions will
remain beyond the reach of direct searches at the LHC. Only part
of the SC and a small fraction of the AF and FP/HB regions will be
explored.  It is therefore interesting to investigate the power of
less direct ways of experimentally testing those regions, including
projected sensitivities, on the most popular constrained SUSY models
like the CMSSM or the NUHM.  In this paper we will investigate two
such observational venues: future measurements of \brbsmumu\ at the
LHC and expected reach of direct search for DM through one-tonne
detectors.

Regarding \brbsmumu, in \cite{Fowlie:2012im} it was also shown that
the impact of the experimental upper bound (at that time) on
\brbsmumu\ was the strongest on the AF region where the SUSY
contribution to the branching ratio is comparable to the SM one, while
the other explored regions were less affected.  A similar conclusion
was reached in\cite{Kowalska:2012gs}, where the impact of the new
positive measurement of \brbsmumu\ was for the first time investigated
in the framework of the Constrained Next-to-Minimal Supersymmetric SM
(CNMSSM), which also features similar favored regions when all the
constraints are simultaneously taken into account.  This points to an
interesting relation between \brbsmumu\ and the relic density
constraint in the AF region.

As stated above, because the current experimental uncertainties are
relatively large, the positive LHCb measurement of \brbsmumu\ is
actually somewhat less constraining for models of new physics
predicting an enhancement of the observable than the previous
exclusion bound.  On the other hand, the systematic and statistical
uncertainties will be greatly reduced when a larger amount of data
comes, and are expected to ultimately achieve the level of 5\%.  It is
therefore interesting to investigate what impact such projected
sensitivities of \brbsmumu\ will have on the favored regions of the
CMSSM and the NUHM.

Our goal is twofold. First, we will present a Bayesian analysis of the
current status of the CMSSM for a much broader range of input
parameters than in\cite{Fowlie:2012im}. We will apply the most recent
experimental determinations of relevant input observables, most
notably the Higgs boson mass and the top quark pole mass, in addition
to the recent positive measurement of \brbsmumu.  We will show that,
in the context of the CMSSM, the expected substantial reduction of the
experimental and theoretical uncertainties in \brbsmumu\ will have the
potential to strongly disfavor basically the whole AF
region. Secondly, we will show that the expected reach of direct
search one-tonne DM detectors will be able to discriminate between the
remaining two favored regions, the SC and the 1TH regions.  On the
other hand, we will show that, unfortunately, a similar conclusion
cannot be reached in the NUHM because of the freedom in adjusting the
pseudoscalar Higgs mass, \mha, and the $\mu$ parameter. Still, in both models one
should be able to distinguish between the SC and the 1TH
regions. Furthermore, any DM signal indicative of the AF region would
strongly disfavor the CMSSM.

Recently, Ref.\cite{Arbey:2012ax} analyzed the impact of the present
measurement of \brbsmumu\ and its future status on random scans of the
CMSSM and on the general MSSM (see also\cite{Altmannshofer:2012ks} for another recent analysis of this 
constraint in the MSSM), showing that a large fraction of the
points generated would be excluded once the projected uncertainties in
\brbsmumu\ are considered.  Our study is partly overlapping but
differs in some important aspects: 1. Our analysis of the CMSSM is
performed as a global Bayesian scan, where the constraints are applied
simultaneously through the likelihood approach (with the exclusion of
XENON100, as explained later). 2. Our main goal is to focus on the
future ability to use \brbsmumu\ to disfavor high probability regions
of models with parameters unified at the scale of grand unification
(GUT).  As a consequence, we do not investigate the general MSSM,
alongside to the CMSSM, but rather the NUHM model.  3. Unlike
in\cite{Arbey:2012ax}, we will also discuss in the detail the
implications of future direct searches of DM.

The paper is organized as follows. In~\refsec{Bsmu} we will
demonstrate semi-analytically how \brbsmumu\ shows a unique discriminating
power over the AF region of the CMSSM. In~\refsec{Method} we will
describe our scanning methodology, and highlight the implementation of
our statistical combination of CMS bounds on SUSY
masses. In~\refsec{Present} we will present our numerical results and
discussion. Finally, we will give our Summary and Conclusions
in~\refsec{Summary}.


\section{\label{Bsmu}\brbsmumu\ in the MSSM}

In this section we first quickly review the analytic form of
\brbsmumu\ in the MSSM and next analyze its implications for the AF
region of the CMSSM and the NUHM.
 
The measurement of the branching ratio is a very good probe of new physics,
since in the SM the decay rate is helicity suppressed, but can get
significant contributions in SUSY.
 
A general expression for the branching ratio
is\cite{Buchalla:1993bv,Misiak:1999yg,Bobeth:2001sq,Bobeth:2001jm} 
\begin{equation}
\textrm{BR}(\bsmumu)=\frac{G_F^2 \alphaem^2 M_{B_s}
  \tau_{B_s}}{16\pi^3}|V_{tb}V_{ts}^{\ast}|^2\sqrt{1-\frac{4
    m_{\mu}^2}{M_{B_s}^2}}\left\{\left(1-\frac{4
      m_{\mu}^2}{M_{B_s}^2}\right)|F_S|^2+|F_P+F_A|^2\right\}\,,\label{bsmumu} 
\end{equation}
where $M_{B_s}$ and $\tau_{B_s}$ are the $B_s$ mass and lifetime, and
$F_A$, $F_P$ and $F_S$ are the axial-vector, pseudo-scalar and scalar
form factor, respectively. In the SM, $F_S$ and $F_P$ are highly
suppressed by helicity conservation, and the only remaining term in
the curly bracket in Eq.~(\ref{bsmumu}) is $|F_A|^2$, where $F_A$
can be expressed in terms of the Wilson coefficient $C_{10}$, the muon
mass $m_{\mu}$, and the $B_s$ decay constant $f_{B_s}$,
$F_A=-im_{\mu}f_{B_s}C_{10}$. The main source of theoretical uncertainty in calculating 
the SM value is the determination of $f_{B_s}$ by the
lattice QCD groups. Ref.\cite{Buras:2010wr} estimates the
$CP$-averaged branching ratio as
$\brbsmumu_{\textrm{SM}}=(3.23\pm0.27)\times 10^{-9}$, while
Ref.\cite{Mahmoudi:2012un} gives a slightly different value,
$\brbsmumu_{\textrm{SM}}=(3.53\pm0.38)\times 10^{-9}$.

Notice that the theoretical calculation should be rescaled by the
effects of $B_s-\bar{B}_s$ oscillations\cite{DeBruyn:2012wk} in order
to be compared with the experimentally measured value.  In this study
we will adopt the value given in\cite{Buras:2010wr} for the
$CP$-averaged SM branching ratio and, following\cite{DeBruyn:2012wk},
take $\brbsmumu_{\textrm{SM}}=3.5\times 10^{-9}$ for the value
rescaled by the effects of $B_s-\bar{B}_s$ oscillations (time
averaged). We differ here from\cite{Arbey:2012ax} where 
$(3.87\pm0.46)\times 10^{-9}$ for the latter was used.

SUSY contributions to $\brbsmumu$ become comparable to the SM when
$F_S$ and $F_P$ are roughly of the same order as $F_A$.  At the leading
order (LO), in the framework of minimal flavor violation, the dominant
SUSY terms in the Wilson coefficients are given by chargino-squark terms
only and are proportional to $\tan^3\beta$\cite{Bobeth:2001sq}.

Following the calculation and notation given in\cite{Bobeth:2001sq}, one can write for $F_P$ and $F_S$
\begin{equation}
F_{S,P}\simeq -\frac{i}{2} M_{B_s}^2 f_{B_s}
C_{S,P}\,,\label{fsfpterms} 
\end{equation}   
where 
\begin{equation}
  C_{S,P}=\mp\frac{m_{\mu}}{4\sin^2\theta_{W}M_W^2}\frac{\tan^3\beta}{\mha^2}\mathcal{F}_{\textrm{LO}}
  \,. \label{cscpterms} 
\end{equation}
The dominant contributions to $\mathcal{F}_{\textrm{LO}}$ is given by the charginos and squarks in the loop,
\begin{eqnarray}
\mathcal{F}_{\textrm{LO}}&\simeq& m_{\chi_1^{\pm}}\sin\theta_U\left\{\sqrt{2} M_W \cos\theta_V \left[-D_3\left(\frac{m_{\tilde{c}_L}^2}{m_{\chi_1^{\pm}}^2}\right)
+D_3\left(\frac{m_{\tilde{t}_1}^2}{m_{\chi_1^{\pm}}^2}\right)\cos^2\theta_t + D_3\left(\frac{m_{\tilde{t}_2}^2}{m_{\chi_1^{\pm}}^2}\right)\sin^2\theta_t\right]\right.\nonumber\\
 &-&\left. m_t \sin\theta_V \sin\theta_t \cos\theta_t \left[D_3\left(\frac{m_{\tilde{t}_1}^2}{m_{\chi_1^{\pm}}^2}\right)-D_3\left(\frac{m_{\tilde{t}_2}^2}{m_{\chi_1^{\pm}}^2}\right)\right]\right\}\label{mchi1term}\nonumber\\
 &+&(\sgnmu)~m_{\chi_2^{\pm}}\cos\theta_U\left\{\sqrt{2} M_W \sin\theta_V \left[D_3\left(\frac{m_{\tilde{c}_L}^2}{m_{\chi_2^{\pm}}^2}\right)-D_3\left(\frac{m_{\tilde{t}_1}^2}{m_{\chi_2^{\pm}}^2}\right)\cos^2\theta_t -D_3\left(\frac{m_{\tilde{t}_2}^2}{m_{\chi_2^{\pm}}^2}\right)\sin^2\theta_t\right]\right.\nonumber\\
 &-&\left. m_t \cos\theta_V \sin\theta_t \cos\theta_t
   \left[D_3\left(\frac{m_{\tilde{t}_1}^2}{m_{\chi_2^{\pm}}^2}\right)-D_3\left(\frac{m_{\tilde{t}_2}^2}{m_{\chi_2^{\pm}}^2}\right)\right]\right\}\,, \label{mchi2term}  
\end{eqnarray}
where we assumed
$\lambda_{22}\equiv V_{cb}V_{cs}^{\ast}/(V_{tb}V_{ts}^{\ast})\simeq-\lambda_{33}=-1$, 
and neglected a term in $\lambda_{11}\equiv V_{ub}V_{us}^{\ast}/(V_{tb}V_{ts}^{\ast})\simeq -10^{-2}$.

The $D_3(x)$ are loop functions,
\begin{equation}
D_3(x)=\frac{x\ln x}{1-x}\,,\label{loopfuncs}
\end{equation}
and $\cos\theta_{U,V}$ and $\sin\theta_{U,V}$ are elements of the
chargino mixing matrices defined such that
$UM_{\chi^{\pm}}V^T=\textrm{diag}(m_{\chi_1^{\pm}},m_{\chi_2^{\pm}})$
(See Appendix B of\cite{Bobeth:2001sq} for notation).\bigskip

For the purpose of this analysis we shall assume that the experimental
uncertainty in $\brbsmumu$ will eventually, with about 50\invfb\ of
data at $\sqrt{s}=14\tev$, be reduced to about 5\%\cite{bsmfutexp}. We
will also assume that the theoretical uncertainty will reach the
precision of 5\%\cite{bsmfutth}.  Hence
$\brbsmumu_{\textrm{proj}}=(3.50\pm0.25)\times10^{-9}$ with both
theoretical and experimental uncertainties added in
quadrature. Further, we will primarily assume that the current SM
value will be confirmed by experimental measurements from LHC with the
above precision, although we will discuss some possible deviations. In
particular, we shall briefly discuss the case that the current LHCb
central value is instead confirmed,
$\brbsmumu_{\textrm{proj}}=(3.20\pm0.23)\times10^{-9}$ and the case
when the assumed ultimate error will be twice as large.

\subsection{Application to the CMSSM}

In the CMSSM, and more generally in unified SUSY models,
Eqs.~(\ref{bsmumu})--(\ref{loopfuncs}) can be greatly simplified
thanks to relations between the different sparticles.

\begin{figure}[t]
\centering
\subfloat[]{%
\label{fig:-a}%
\includegraphics[width=0.45\textwidth]{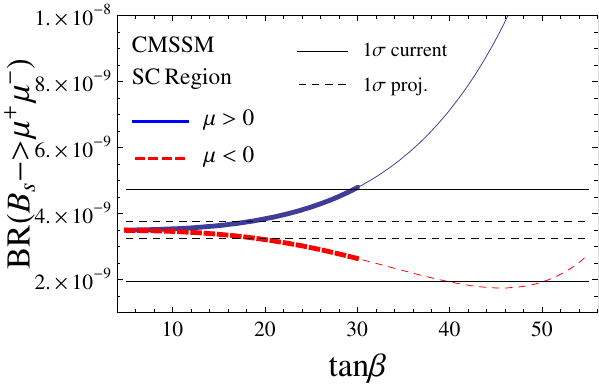}
}%
\subfloat[]{%
\label{fig:-b}%
\includegraphics[width=0.45\textwidth]{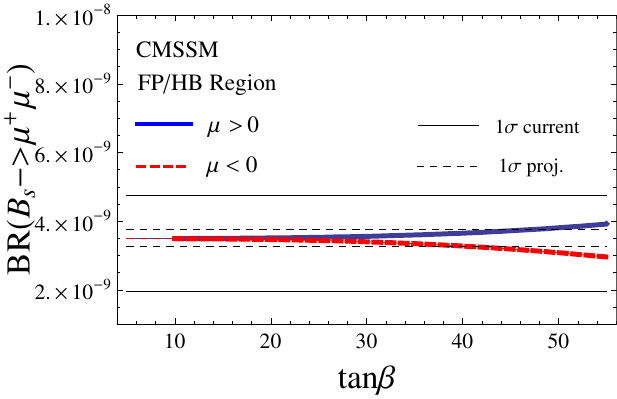}
}%
\hspace{1pt}%
\subfloat[]{%
\label{fig:-c}%
\includegraphics[width=0.45\textwidth]{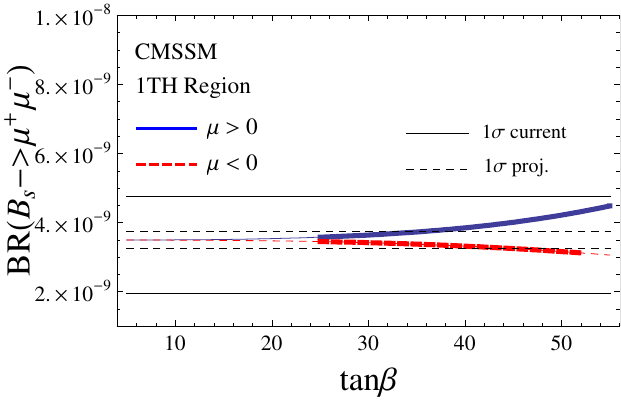}
}%
\subfloat[]{%
\label{fig:-d}%
\includegraphics[width=0.45\textwidth]{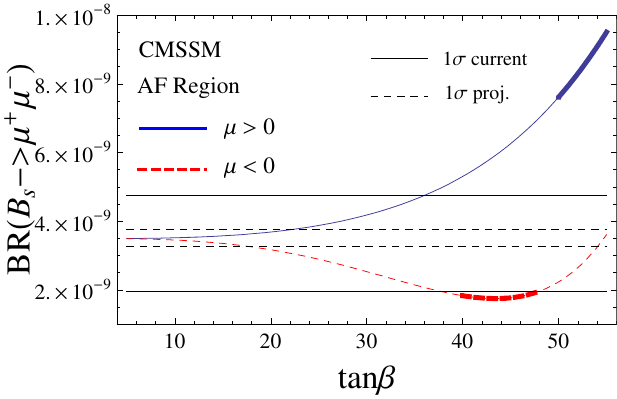}
}%
\caption[]{ The dependence of \brbsmumu\ on \tanb\ in the CMSSM. Solid
  blue line: $\mu>0$; dashed red line: $\mu<0$. Solid horizontal
  lines: current $1\sigma$ error on $\brbsmumu$; dashed horizontal
  lines: projected error. The thick lines show the values of \tanb\ typical of
  each region. \subref{fig:-a} SC region. \subref{fig:-b} FP/HB
  region. \subref{fig:-c} 1TH region. \subref{fig:-d} AF
  region.}
\label{fig:bsmumu}
\end{figure} 

The first and the third line in the right-hand side of
Eq.~(\ref{mchi1term}) are always opposite in sign and cancel each
other out to a good approximation.  Since in the SC and AF regions
the neutralino is strongly
bino-dominated, it follows that $m_{\chi_1^{\pm}}\simeq M_2$ and
$m_{\chi_2^{\pm}}\simeq \mu$, so that
$\sin\theta_U\sin\theta_V\simeq0$ and the fourth line in
Eq.~(\ref{mchi1term}) is dominant.  In the FP/HB and the 1TH  regions, where the
roles of $m_{\chi_1^{\pm}}$ and $m_{\chi_2^{\pm}}$ are interchanged,
$\cos\theta_U\cos\theta_V\simeq0$ so that the second term is thus
dominant.  In all of the favored regions, by remembering that
$\sin\theta_t\cos\theta_t\simeq m_t
A_t/(m_{\tilde{t}_1}^2-m_{\tilde{t}_2}^2)$ and that, for moderate to
large \tanb, a change in \sgnmu\ implies a change in the sign of
$\sin\theta_U$, as $\tan\theta_U\propto 1/(\cos\beta
M_2+\sin\beta\mu)$, one can recast Eq.~(\ref{mchi1term}) as
  
\begin{equation}
\mathcal{F}_{\textrm{LO}}\simeq -\mu \mathcal{D}_3 \frac{m_t^2
  A_t}{m_{\tilde{t}_1}^2-m_{\tilde{t}_2}^2}\,,\label{simplyform} 
\end{equation}
where $\mathcal{D}_3$ is given by differences of $D_3$ functions and is in general of order 0.1--0.3.

In our numerical analysis we will use full calculations to higher
order, given by the latest numerical codes, but we can use the above
approximation to show how a projected better determination of \brbsmumu\
can affect the four regions favored by the correct relic density, as
explained above.
  
For choices of parameters typical of the SC region,
$\mathcal{F}_{\textrm{LO}}$ is the largest, driven up by large values
of $\mu$, $\mu>1000\gev$, and by maximal $A_t/\msusy$, which gives
$\mathcal{D}_3\simeq 0.3$.  On the other hand, as can be seen from
Eq.~(\ref{cscpterms}), the branching ratio is suppressed by moderate
\tanb\ values, $\tanb\sim 5-30$ typical for the SC region.  The
\tanb\ dependence of \brbsmumu\ for a point representative of the SC
region ($\mzero=226\gev$, $\mhalf=827\gev$, $\azero=-1375\gev$) is
shown in Fig.~\ref{fig:bsmumu}\subref{fig:-a}. The solid blue line
gives the case $\mu>0$ and the dashed red line the case $\mu<0$.  The
thick lines show values of \tanb\ characteristic of the SC
region. The solid horizontal lines give the current $1\sigma$
theoretical and experimental uncertainty on the measurement added in
quadrature.  The dashed horizontal lines denote our estimated $1\sigma$
projected uncertainties added in quadrature.
 
In the FP/HB region and, at large mass parameters, in the 1TH region,
$\mathcal{F}_{\textrm{LO}}$ is the smallest, since $A_t/\msusy$ is
minimal and $m_{\chi_1^{\pm}}\approx\mu\lesssim 1\tev$.  Moreover, the
branching ratio is suppressed by large \mha, even if \tanb\ can assume
a wide range of values. We show in
Fig.~\ref{fig:bsmumu}\subref{fig:-b} the \tanb\ dependence of
\brbsmumu\ for a point representative of the FP/HB region
($\mzero=3447\gev$, $\mhalf=866\gev$, $\azero=730\gev$), and in
Fig.~\ref{fig:bsmumu}\subref{fig:-c}, the same for a point
representative of the $1\tev$ higgsino region ($\mzero=7989\gev$,
$\mhalf=2854\gev$, $\azero=-767\gev$).  The color code is the same as
in Fig.~\ref{fig:bsmumu}\subref{fig:-a}.

Finally, but most importantly for the purpose of this paper, in the AF
region \tanb\ \textit{has to be large} in order to yield the correct
$\abundchi$, as we will explain in the following subsection.  Thus,
the measured value of \brbsmumu\ becomes important in constraining the
parameter space ($\mu$ and \mha\ are comparable to the SC region,
while stop mixing is not as large, so that $\mathcal{D}_3\sim
0.15$--0.2).  In Fig.~\ref{fig:bsmumu}\subref{fig:-d} we again
indicate with thick lines the ranges of \tanb, for both signs of
$\mu$, which give \abundchi\ within $1\sigma$ (theoretical +
experimental uncertainties added in quadrature) of the central
value. The difference in the allowed values of \tanb\ for different
\sgnmu\ is a feature of the AF region, and was already observed
in\cite{Fowlie:2012im}. We will explain this in the next
subsection. Notice also that, for $\mu<0$ the calculated value of
\brbsmumu\ is more than $1\sigma$ \textit{below} the SM value, since
the form factors $F_P$ and $F_A$ undergo destructive interference, and
one is left with a small value of $F_S$.

\subsection{\label{Afunreg} The $A$-funnel region vs \brbsmumu}

The AF region is particularly sensitive to the determination of
\brbsmumu\ because at the LO the relic density there depends mainly on
the same parameters, \mha\ and \tanb; see Eq.~(\ref{cscpterms}).

AF annihilation occurs when the mass of the pseudoscalar $A$ is close
$2\mchi$, so the lightest neutralino can efficiently annihilate into
SM fermions (mostly $b$-quarks) through the $s$-channel exchange of
$A$.  For an almost pure bino LSP (higgsino components of order
$10^{-2}$ or less) and \tanb\ in the range 20--60 (for lower values
the channel $\chi\chi\rightarrow Z h$ becomes dominant) one
obtains\cite{Drees:1992am}
\begin{equation}
\Omega_{\chi}h^2\approx \frac{3\times10^{-27}\textrm{cm}^3/\textrm{s}}{\langle\sigma v\rangle}\,,
\label{Omega}
\end{equation}
where 
\begin{equation}
\sigma v\approx\frac{\textrm{const}}{\mchi^2}\frac{\tan^2\beta}{(4-\mha^2/\mchi^2)^2+(\Gamma_A\mha/\mchi^2)^2}\left(1+\frac{v^2}{4}\right)\,,\label{sigmav}
\end{equation}
with the $A$ width being, for $\mu>0$ ($\mu<0$), $\Gamma_A\approx 1.3~(2.0)\times 10^{-5}~m_A\tan^2\beta$. 
The constant in Eq.~(\ref{sigmav}) depends moderately on kinematical factors, on the neutralino composition, and on \tanb. For masses given in \gev, 
its value is 
$\sim 10^{-25}$ cm$^3 \gev^2/$s. 
The correct relic density is generally achieved for a difference $|\ma-2\mchi|$ not exceeding 100\gev\cite{Roszkowski:2001sb}.

In the CMSSM with bino-like DM, \mha\ can in principle be close to $2
\mchi$ for wide ranges of \mzero\ and \mhalf, if the value of \tanb\
is properly adjusted: the mass of a bino-like LSP is approximately
given by $\mchi\approx 0.44~\mhalf$ while $\mha\sim\kappa~\mhalf\times
f\left(\frac{50}{\tanb}\right)$, where $\kappa$ is of order 0.8--0.9, $f(x)$ is a monotonically
increasing function of $x$, and $f(1)=1$.  As a consequence, for large
\tanb, when \mhalf\ increases $2 \mchi$ increases faster than
\mha. Hence, in order to get the resonance for larger \mhalf, one
needs to assume smaller \tanb.  However, this does not mean that the
correct relic density can always be obtained since, as
Eq.~(\ref{sigmav}) shows, even for $\mha\simeq2\mchi$ the cross
section becomes suppressed with increasing neutralino mass for any
given \tanb.

%
\begin{figure}[t]
\centering
\subfloat[]{%
\label{fig:-a}%
\includegraphics[width=0.45\textwidth]{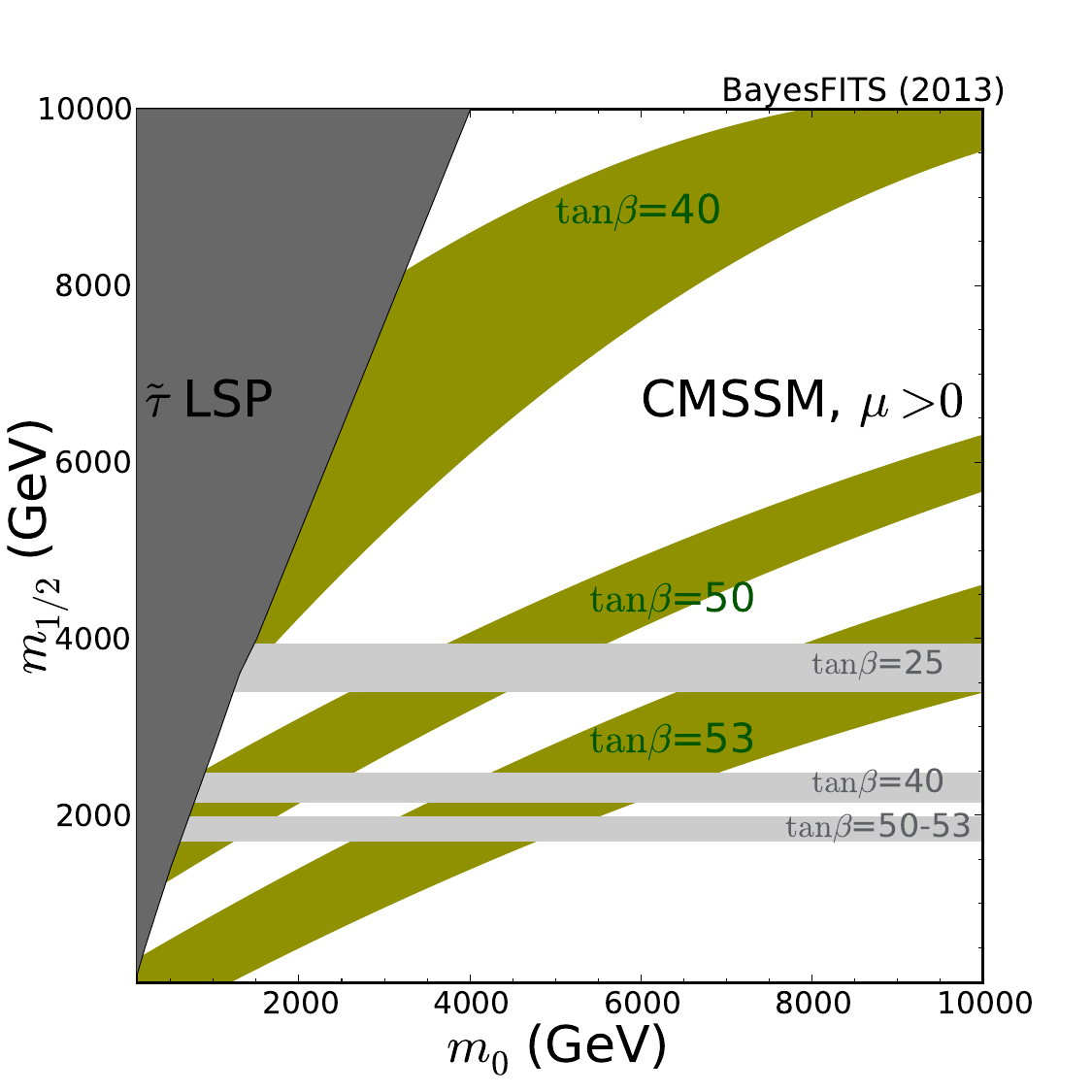}
}%
\subfloat[]{%
\label{fig:-b}%
\includegraphics[width=0.45\textwidth]{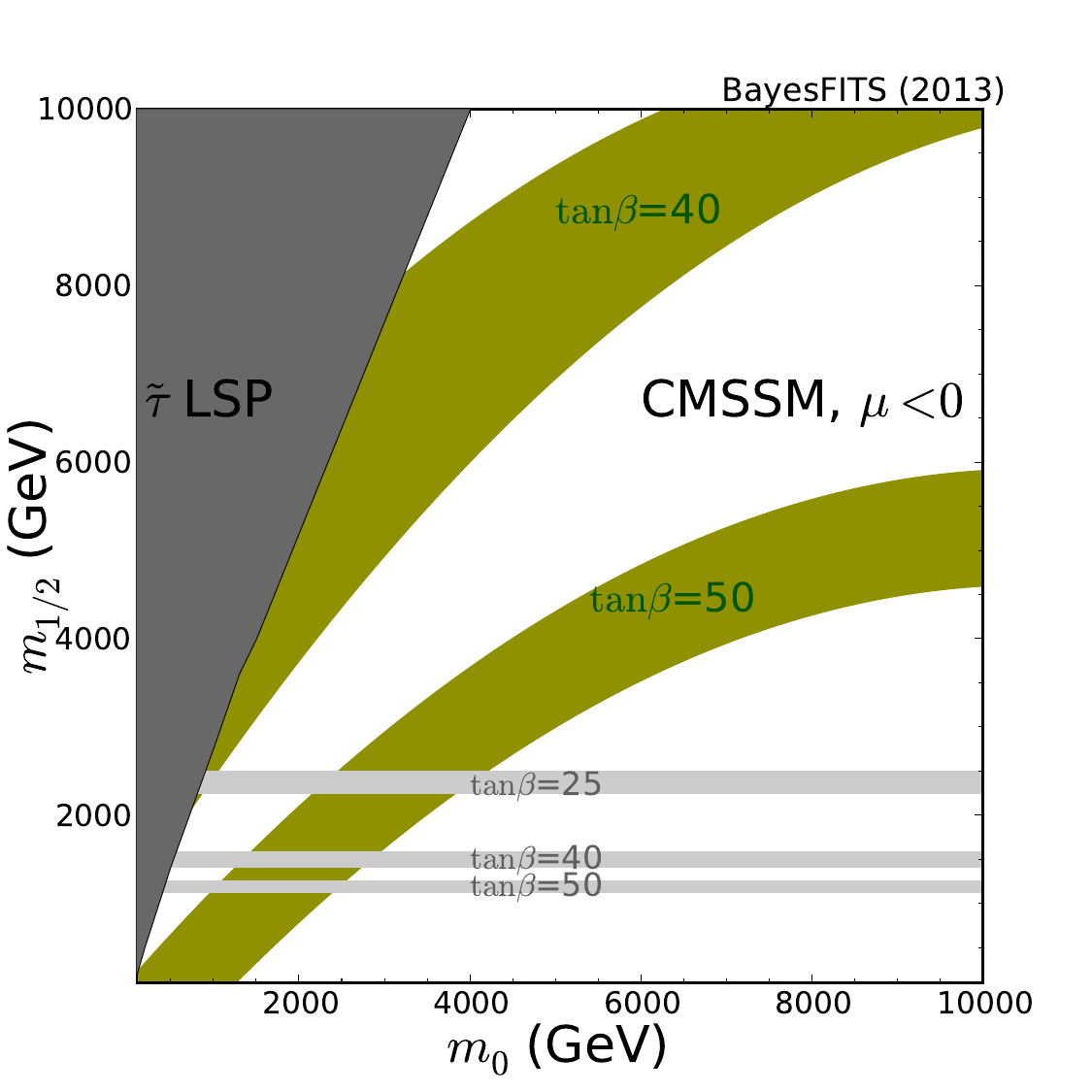}
}%
\caption[]{Green bands: the regions of the CMSSM where $\mha=2\mchi$
  for $-10\tev\leq\azero\leq 10\tev$.  Light gray bands: the regions
  where $\sigma v$ for bino-like DM gives the correct $\abundchi$
  through $A$-resonant annihilation.  Dark gray: the neutralino is not
  the LSP. \subref{fig:-a} $\mu>0$. \subref{fig:-b} $\mu<0$.}
\label{fig:AFres}
\end{figure} 

In Fig.~\ref{fig:AFres} the green bands show the regions of the
(\mzero, \mhalf) plane over which the condition $\mha=2\mchi$ is
satisfied for fixed \tanb\ and $-10\tev\leq\azero\leq
10\tev$. Figure~\ref{fig:AFres}\subref{fig:-a} shows the case $\mu>0$
and Fig.~\ref{fig:AFres}\subref{fig:-b} $\mu<0$.  Note that, this is
achieved when, for the same \tanb, \mhalf\ is slightly smaller for
negative $\mu$ than for positive $\mu$.  The reason lies in a one-loop
tadpole contribution to the effective potential of the
model\cite{Barger:1993gh}.  The corrections due to sfermions,
charginos and neutralinos explicitly depend on \sgnmu, leading to
positive (negative) contribution to $\mha^2$ for $\mu>0$ ($\mu<0$).
Therefore, when the other parameters of the model are left unchanged,
\mha\ is slightly smaller for negative $\mu$.
For the same reason, values of \tanb\ larger than 50 cannot be obtained for 
negative $\mu$ since they would lead to $\mha^2<0$ and no electro-weak symmetry breaking. 

Using Eqs.~(\ref{Omega}) and (\ref{sigmav}), one can now calculate the
ranges of \mhalf\ that for a given \tanb\ would allow to obtain the
correct relic density (within $1\sigma$ of the experimental central
value) when $\mha=2\mchi$.  We show them in
Figs.~\ref{fig:AFres}\subref{fig:-a} and
\ref{fig:AFres}\subref{fig:-b} as gray horizontal stripes. One can see
that the AF region of the CMSSM is \textit{confined} to a relatively
small part of the (\mzero, \mhalf) plane, where the green and the grey
bands of the same \tanb\ intersect. This is also true in the more
realistic case where \mha\ and $2\mchi$ are within 100\gev\ from one
another, as is confirmed by  numerical scans.  As a consequence,
\tanb\ is also constrained in the AF region: it can take values in the
range 48--55 for positive $\mu$, and 38--50 for negative $\mu$.  This
is the ranges we highlighted in boldface in
Fig.~\ref{fig:bsmumu}\subref{fig:-d}.

One can also see in Fig.~\ref{fig:bsmumu}\subref{fig:-d} that the values
of \mha\ and \tanb\ typical of the AF region are the ones that
show most tension with the current measurement of
\brbsmumu. Moreover, it is clear that future, more precise, measurements of the
branching ratio will have the potential to exclude a broad range of the
(\mzero, \mhalf) parameter space corresponding to the AF region.

In the NUHM the situation is quite different. The additional soft mass
parameters in the Higgs sector \mhu\ and \mhd\ can be traded, through conditions of
electroweak symmetry breaking (see, \eg,\cite{Roszkowski:2009sm}), for
\mha\ and $\mu$:
\bea
\label{ewsb1:eq}
\mu^2&=& \frac{\mhd^2 - \mhu^2\, \tan^2\beta}{\tan^2\beta -1} -\frac{1}{2}\mz^2,\\
\mha^2 &=& \mhd^2 + \mhu^2+2\mu^2,
\label{ewsb2:eq}
\eea
which can be adjusted to satisfy the resonance condition,
independently of \tanb, for much wider ranges of both \mzero\ and
\mhalf.  Therefore, in the NUHM the AF region giving the correct relic
density is not as well localized and occupies a wide part of the
parameter space.  Also \tanb\ is now allowed to assume a much wider
range of input values, which is crucial from the point of view of
satisfying the \brbsmumu\ constraint. We will come back to this point later.


\section{\label{Method}Scanning Methodology and Constraints}


In order to examine the impact of the most recent constraints,
including \brbsmumu, on the parameter space of the CMSSM and the NUHM
we use the Bayesian approach.  We follow the procedure outlined in
detail in Refs.\cite{Fowlie:2011mb,Roszkowski:2012uf,Fowlie:2012im}.  Our goal
is to map out the 68\% and 95\% credible regions of $p(m|d)$, the
posterior probability density function (pdf), given by Bayes' theorem,
\be
p(m|d)=\frac{p(d|\xi(m))\pi(m)}{p(d)}\,, 
\label{eq:1}
\ee
where $p(d|\xi(m))\equiv\mathcal{L}$ is the likelihood function, which
describes the probability of obtaining the data $d$ given the computed
value of some observable $\xi(m)$, which is a function of the model's
parameters $m$. $\mathcal{L}$ incorporates the information about the constraints, 
as well as their experimental and theoretical uncertainties. Prior probability
$\pi(m)$ encodes assumed range and distribution of $m$.  Finally,
$p(d)$ is the evidence and is a normalization constant as long as
only one model is considered, but serves as a comparative measure for
different models or scenarios.

Bayes' theorem provides an efficient and natural procedure for drawing
inferences on a subset of $r$ specific model parameters (including
nuisance parameters), or observables, or a combination of both, which
we collectively denote by $\psi_i$.  They can be obtained through
marginalization of the full posterior pdf, carried out as
\be
p(\psi_{i=1,..,r}|d)=\int p(m|d)d^{n-r}m\,,\label{marginalize}
\ee
where $n$ is the total number of input parameters.


\subsection{Experimental Constraints}

\begin{table}[t]\footnotesize
\begin{center}
\begin{tabular}{|l|l|l|l|l|l|}
\hline
Measurement & Mean or Range & Error:~(Exp.,~Th.) & Distribution & Ref.\\
\hline
Combination of: & & & & \\
\razorfourfb, \seven & See text 	& See text 	& Poisson &\cite{Chatrchyan:2012uea}\\ 
\alphaTelefb, \eight & See text 	& See text 	& Poisson &\cite{newalphat}\\ 
\hline 
\mhl\ by CMS & $125.8\gev$ & $0.6\gev, 3\gev$ & Gaussian &\cite{cmshiggs} \\
\abundchi 			& $0.1120$ 	& $0.0056$,~$10\%$ 		& Gaussian &  \cite{Komatsu:2010fb}\\
\deltagmtwomususy $\times 10^{10}$ 	& $28.7 $  	& $8.0$,~$1.0$ 		& Gaussian &  \cite{Bennett:2006fi,Miller:2007kk} \\
\brbxsgamma $\times 10^{4}$ 		& $3.43$   	& $0.22$,~$0.21$ 		& Gaussian &  \cite{bsgamma}\\
\brbutaunu $\times 10^{4}$          & $1.66$  	& $0.33$,~$0.38$ 		& Gaussian &  \cite{Amhis:2012bh}\\
$\Delta M_{B_s}$ & $17.719\ps^{-1}$ & $0.043\ps^{-1},~2.400\ps^{-1}$ & Gaussian & \cite{Beringer:1900zz}\\
\sinsqeff 			& $0.23116$     & $0.00012$, $0.00015$             & Gaussian &  \cite{Beringer:1900zz}\\
$M_W$                     	& $80.385$      & $0.015$, $0.015$               & Gaussian &  \cite{Beringer:1900zz}\\
\hline
$\brbsmumu_{\textrm{current}}\times 10^9$			& 3.2
&  $+1.5-1.2$, 10\% (0.32) & Gaussian &  \cite{Aaij:2012ct}\\
$\brbsmumu_{\textrm{proj}}\times 10^9$			& 3.5~($3.2^{\ast}$)  	&  0.18 ($0.16^{\ast}$), 5\%  [0.18 ($0.16^{\ast}$)]      & Gaussian &  \cite{Aaij:2012ct}\\
\hline 
\end{tabular}

\footnotesize{$\ast$ We will also consider the case of projected uncertainties around the current measured central value.}
\caption{The experimental constraints that we apply to
  constrain model parameters. } 
\label{tab:exp_constraints}
\end{center}
\end{table}

The central object in our analysis is the likelihood
function as the place where theoretical predictions are compared
with experimental data.
The constraints that we include in the current analysis are
listed in Table~\ref{tab:exp_constraints}. 
As a rule, following the procedure developed
earlier\cite{deAustri:2006pe}, we implemented positive measurements through
a Gaussian likelihood, in which the experimental and theoretical uncertainties were added in quadrature.
For the Higgs mass, we used the most recent CMS determination of its central value and experimental uncertainty, 
as it is in perfect agreement with the determination obtained by ATLAS at the end of the \eight\ run.
The theoretical uncertainty was estimated to be 3\gev\cite{Fowlie:2012im,Heinemeyer:2011aa}.

As stated above, for \brbsmumu\ we considered two cases: 

1. The \textit{current} measurement at LHCb, for which we adopted a
theoretical uncertainty of 10\% of the measured value (see
next-to-bottom row in Table~\ref{tab:exp_constraints}), in agreement
with\cite{Mahmoudi:2012un} once the  uncertainty due to the
top pole mass ($\sim1\%$) is subtracted. We do so because in our scans
the top mass is one of the nuisance parameters and the effect of
varying it is included parametrically.

2. The \textit{projected} `best-case' scenario for the determination
of \brbsmumu, where the experimental and theoretical uncertainties are
both reduced to 5\% of the measured value (see bottom row in
Table~\ref{tab:exp_constraints}), as explained in Sec.~\ref{Bsmu}.  In
addition, as a sensitivity test, we considered both the case where the
measurement will be narrowed down to the time-averaged SM value,
$3.5\times10^{-9}$, and the case where the current central LHCb
experimental value, $3.2\times10^{-9}$, will be confirmed by future
sensitivities.  This second case can in principle improve the fit for
the AF region in the $\mu<0$ case, since the branching ratio there
assumes values more than $1\sigma$ below the SM determination (see
Fig.~\ref{fig:bsmumu}\subref{fig:-d}
and\cite{Fowlie:2012im}). Finally, we will double the assumed error
around the SM value, again as a sensitivity test.
  
Following the procedure already adopted in our previous papers, we did
not include the XENON100 upper bound explicitly in the likelihood
function.  The theory uncertainties are very large (up to a factor of
10) and strongly affect the impact of the experimental limit on the
parameter space.  The main source of error (the so-called $\Sigma_{\pi
  N}$ term\cite{Ellis:2008hf}) arises from different, and in fact
partly incompatible, results following from different calculations
based on different assumptions and methodologies.  Such uncertainties
do not follow a particular statistical distribution, and are not well
suited for inclusion in a likelihood function.  Moreover, we showed in
a previous publication\cite{Roszkowski:2012uf} that, when smearing out
the XENON100 limit with a theoretical uncertainty of order ten times
the given value of \sigsip\ the effect on the posterior is negligible
for regions of parameter that appear up to one order of magnitude above
(and below) the experimental limit.  However, even if we do not
include the XENON100 bound in the likelihood, below we shall comment
on its possible effects on the posterior pdf.
    
The likelihood for limits from direct SUSY searches deserves a more
detailed explanation, which we give in the following subsection.

\subsubsection{\label{combo} Combination of CMS SUSY search limits}

In previous work\cite{Fowlie:2011mb,Fowlie:2012im} we presented a methodology 
for deriving approximate but accurate likelihood functions for two of the direct SUSY searches 
with all-hadronic final states at CMS:
$\alpha_T$ (Ref.\cite{Fowlie:2011mb}) and razor (Ref.\cite{Fowlie:2012im}).
Our approximation correctly reproduced the 
95\%~CL exclusion bounds of those searches in the (\mzero, \mhalf) plane.
In\cite{Kowalska:2012gs} we then showed that the same procedure for the razor search could be extended to the CNMSSM. 

The likelihood maps were developed through a step-by-step procedure which included
generation of the SUSY signal at the scattering level with
PYTHIA6.4\cite{PYTHIA} and a simulation of the CMS detector response
with PGS4\cite{PGS4} to calculate the efficiency once the kinematic
cuts were applied.  The obtained signal yields were finally
statistically compared to the publicly available observed and
background yields of the searches to construct the likelihood map.
 
As mentioned in the Introduction, the most constraining limit for the
CMSSM presently comes from the ATLAS search for squarks and gluinos
with jets and missing transverse energy in the final states, with
5.8\invfb\ of data at \eight\cite{ATLAS-CONF-2012-109}.  The recent
limits produced by the CMS Collaboration with comparable or larger
luminosity\cite{newalphat} are slightly
weaker.  On the other hand, riding on our accurate method for
constructing the likelihood function for all-hadronic SUSY search
limits with the information provided by the CMS Collaboration, we are
in a position of deriving an approximate statistical combination of
the CMS searches at \seven\ and \eight.

We prefer to follow this procedure rather than taking the ATLAS limit
as a hard cut (as recently done, \eg, in\cite{CahillRowley:2012kx,
Strege:2012bt}) for one important reason.  Most recent analyses of
the CMSSM have pointed out that the region of parameter space which
provides the best fit to the constraints (particularly the Higgs mass) is the SC
region. Since this is the region directly adjacent to the exclusion
bounds, accurate modelling of the likelihood function becomes important.

In what follows we briefly summarize the methodology adopted for the
razor in our previous papers, since it will be used again here.  We then
proceed to statistically combining it with the most recent
CMS \alphaT\ search to update our exclusion bound.

\paragraph{Razor 4.4\invfb, \seven}\mbox{}\\
The CMS razor search, based on 4.4\invfb\ of \seven\ data, found no
excess of events over the SM prediction.  In deriving the likelihood
map for the razor analysis we followed the CMS procedure described
in\cite{Chatrchyan:2012uea}.  All accepted events were divided into 38
separate bins in the two-dimensional space of the razor variables
$R^2$ and $M_R$, and the likelihood of observing a certain number
events in a given bin was defined as a Poisson distribution convolved
with a Gaussian or log-normal function that would take care of the
predicted error on the background yields.  The details of our analysis
can be found in\cite{Fowlie:2012im}.

\begin{figure}[t]
\centering
\subfloat[]{%
\label{fig:-a}%
\includegraphics[width=0.45\textwidth]{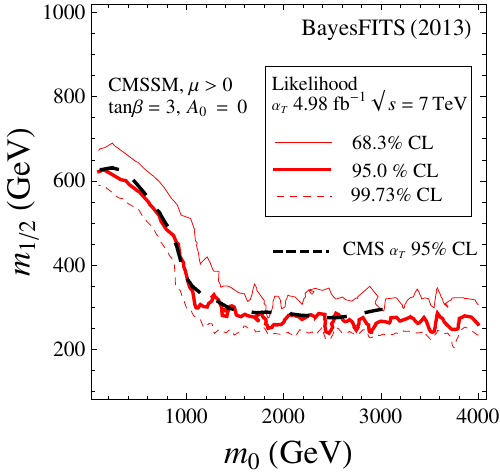}
}%
\subfloat[]{%
\label{fig:-b}%
\includegraphics[width=0.45\textwidth]{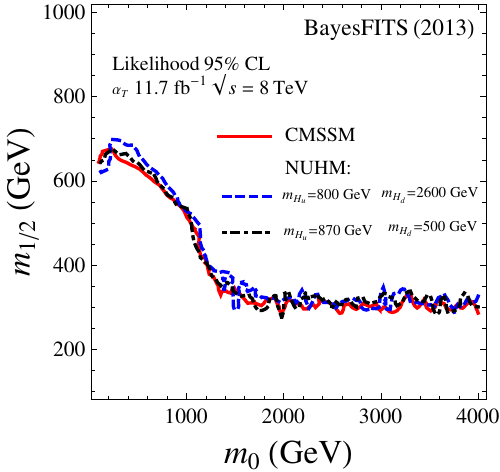}
}%
\hspace{1pt}%
\subfloat[]{%
\label{fig:-c}%
\includegraphics[width=0.45\textwidth]{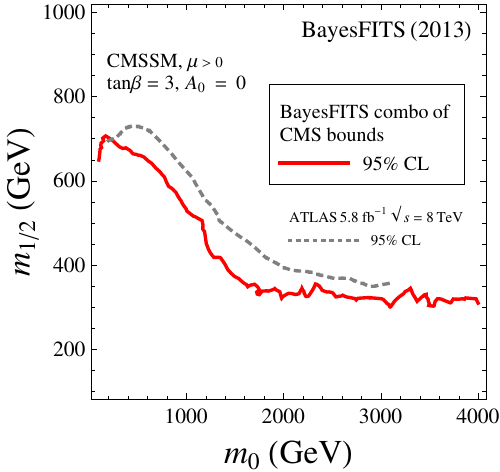}
}%
\caption[]{\subref{fig:-a} The 68.3\%~CL (red solid thin), 95.0\%~CL
  (red solid thick), and 99.7\%~CL (red dashed thin) exclusion bounds
  for the CMSSM from our approximation of the $\alpha_T$ likelihood
  ($\sqrt{s}=7\tev$, $\sim5\invfb$) compared to the original CMS
  95\%~CL exclusion bound (dashed black). \subref{fig:-b} 95\%~CL
  exclusion bound (solid red) for the CMSSM from our approximation of
  the $\alpha_T$ likelihood ($\sqrt{s}=8\tev$, $\sim12\invfb$)
  compared with the bounds obtained for the NUHM, when $\mhu<\mhd$
  (dashed blue) and $\mhd<\mhu$ (dot-dashed black). \subref{fig:-c}
  95\%~CL exclusion bound (solid red) for the CMSSM from our
  combination of CMS searches (solid red) compared to the current ATLAS
  bound (dotted gray).}
\label{fig:CMScombination}
\end{figure} 

\paragraph{\alphaT\ 11.7\invfb, \eight}\mbox{}\\
The CMS \alphaT\ search, performed with 11.7\invfb\ of data based on
\eight\ $pp$ collisions, shows no significant deviation from the SM
prediction\cite{newalphat}. In deriving the likelihood
map we followed closely the CMS procedure and our methodology
presented
in\cite{Fowlie:2011mb,Roszkowski:2012uf}. 
The accepted events were divided into 8 separate boxes, according to
the number of jets originating from $b$-quarks, $n_b=0,1,2,3$ or $\ge4$ and to the 
number of reconstructed jets per event, $2\le n_j\le 3$ and
$n_j\ge 4$. In every box, the events were classified based on the value of the
variable $H_T$, defined as the sum of all jets' transverse energies.
The likelihood for observing $o_i$ events in the $i$-th bin, given the
known number of the expected events $s_i$, and the number of the
expected SM background events $b_i$, is given by a Poisson
distribution convolved with a Gaussian, to account for the predicted
error on the background yield.  The ranges of $H_T$ in every bin,
together with the corresponding numbers of the observed events,
expected background events, and errors on the expected background
yield provided by the CMS Collaboration, are given
in\cite{alphatsite}.

Since the CMS Collaboration has not provided for this search the official 95\%~\cl\ exclusion bound in the CMSSM, 
we validated our likelihood map procedure for the \alphaT\ analysis with the official \alphaT\ 4.98\invfb, \seven\ contour
given in\cite{Chatrchyan:2012wa}. In Fig.~\ref{fig:CMScombination}\subref{fig:-a} we show the comparison between our simulation and the official plot. 
As one can see, we obtained very good agreement.
In Fig.~\ref{fig:CMScombination}\subref{fig:-b} we show our 95\%~CL contour for the \alphaT\ 11.7\invfb, \eight\ as a
solid red line. It is a big
advantage of the likelihood map methodology that it allows one to derive
likelihood functions for SUSY searches even where the official limits
are not available.
\medskip

Furthermore, we also show in
Fig.~\ref{fig:CMScombination}\subref{fig:-b} that the derived
exclusion limit can be applied not only to the CMSSM, but also to the
NUHM. The exclusion bounds obtained for two different choices of the
parameters \mhu\ and \mhd\ (shown in dashed blue and dot-dashed black)
do not differ from the CMSSM one. The reason is that the soft masses
of the Higgs sector enter the one-loop renormalization group equations
of the first two generation quarks only by the terms multiplied by the
Yukawa couplings, and therefore are strongly suppressed, while the term
proportional to the difference $(\mhu^2-\mhd^2)$ is multiplied by the
factor $g_1^2/10$ and is also negligible, unless the mass difference
is very large. The NUHM exclusion limits shown at
Fig.~\ref{fig:CMScombination}\subref{fig:-b} correspond precisely to
the choice of parameters that would maximize the difference
$|\mhu^2-\mhd^2|$, and at the same time remain in agreement with the
physicality condition.

In\cite{Fowlie:2012im} we showed that the 95\%~\cl\ limit based on the
4.4\invfb\ razor search is not affected by the change of the sign of
parameter $\mu$.  The same is true for the NUHM.

\paragraph{Limit combination procedure and results}\mbox{}\\
In our approximate combination of the recent SUSY searches by CMS, we
used all bins considered in the CMS \alphaT\ 11.7\invfb, \eight\ analysis\cite{newalphat}, as well as the ones from
the razor 4.4\invfb, \seven\cite{Fowlie:2012im}. Following the statistical
approach of Modified Frequentist Confidence Levels\cite{Junk:1999kv}
we assumed that the two searches are statistically independent (since they are based on different data sets) and
we treated every bin as a statistically independent counting
experiment.  Then the combined likelihood is a product of the
likelihoods for the two separate searches.
The results of such an approximation are presented in
Fig.~\ref{fig:CMScombination}\subref{fig:-c}, which shows a comparison
of the 95\%~\cl\ lines for the 5.8\invfb\ ATLAS search at
$\sqrt{s}=8\tev$ (dotted gray) and our combination of CMS results
described above (solid red).




\subsection{Scanning tools and parameter ranges}

In this analysis we used the package BayesFITS which calls several
external, publicly available tools: for sampling it uses
MultiNest\cite{Feroz:2008xx} with evidence tolerance factor set to
0.5, sampling efficiency equal to 0.8, and number of live points equal
to 4000 (CMSSM) or 10000 (NUHM).

\begin{table}[t]\footnotesize
\begin{tabular}{|l|l|l|l|}
\hline 
CMSSM parameter & Description & Prior Range & Prior Distribution \\
\hline 
\mzero        	& Universal scalar mass          & 0.1, 20 	& Log\\
\mhalf		& Universal gaugino mass         & 0.1, 10 	& Log\\
\azero        	& Universal trilinear coupling   & -20, 20	& Linear\\
\tanb	        & Ratio of Higgs vevs            & 3, 62 	& Linear\\
\sgnmu		& Sign of Higgs parameter        & +1 or $-1$ 		& Fixed\\
\hline 
Nuisance & Description & Central value $\pm$ std. dev. & Prior Distribution \\
\hline 
\mt           	& Top quark pole mass 	& $173.5\pm1.0$ 	& Gaussian\\
\mbmbsmmsbar 	& Bottom quark mass	& $4.18\pm0.03$ 	& Gaussian\\
\alphasmzms	& Strong coupling	& $0.1184\pm0.0007$   	& Gaussian\\
1/\alphaemmz 	& Reciprocal of electromagnetic coupling  	& $127.916\pm 0.015$ 	& Gaussian\\
\hline 
\end{tabular}
\caption{Priors for the parameters of the CMSSM  and for the SM nuisance
  parameters used in our scans. Soft masses and \azero\ are in
  TeV. Top quark pole mass and bottom quark mass are in GeV.}
\label{tab:priorsCMSSM}
\end{table}

Mass spectra were computed with \softsusy\ v3.3.6\cite{softsusy} and
passed via SUSY Les Houches Accord format to \superiso\
v3.3\cite{superiso} to calculate \brbxsgamma, \brbsmumu, \brbutaunu,
and \deltagmtwomususy. \delmbs, \sinsqeff\ and $M_W$ are calculated
with FeynHiggs\cite{feynhiggs:00}.  DM observables, such as the relic
density and direct detection cross sections, are calculated with
\micromegas\ 2.4.5\cite{micromegas}.

The prior ranges and metric adopted for scanning the CMSSM and
nuisance parameters are given in~\reftable{tab:priorsCMSSM}.  We only
scanned in log priors for the mass parameters, as it was proven in
many previous studies\cite{arXiv:0809.3792} that flat priors in the
CMSSM unduly favor the large-scale regions of the parameter space
(volume effect). Moreover, the correlation between \brbsmumu\ and the
AF region, which we expose in this study, becomes unobservable once
large values of \ma\ become favored by the scan.

Note that, compared to\cite{Fowlie:2012im}, we significantly extended
the ranges of \mzero, \mhalf\ and \azero.  We performed our scans for
$\mu>0$ and $\mu<0$ separately. For negative $\mu$ we did not include
the \gmtwo\ constraint, since its only effect would be to worsen the
overall fit (see\cite{Fowlie:2012im} for a detailed discussion of this
issue), while the observable is very poorly fit anyway.

The prior ranges and metric for the NUHM parameters are given in~\reftable{tab:priorsNUHM}. 
We performed several  scans with different choices of ranges and priors.
As we will explain in more detail in the next section, we selected the ranges that allowed us to 
most strongly see the possible correlation between \brbsmumu\ and the AF region.

\begin{table}[t]\footnotesize
\begin{centering}
\begin{tabular}{|l|l|l|l|}
\hline 
NUHM parameter & Description & Prior Range & Prior Distribution \\
\hline 
\mzero        	& Universal scalar mass          & 0.1, 4 (0.1, $20^{\ast}$)	& Log (Linear)\\
\mhalf		& Universal gaugino mass         & 0.1, 4 (0.1, 10)	& Log (Linear)\\
\azero        	& Universal trilinear coupling   & -7, 7 (-20, 20) & Linear\\
\tanb	        & Ratio of Higgs vevs            & 15, 35 (3, 62) & Linear\\
\sgnmu		& Sign of Higgs parameter        & +1 or $-1$ 		& Fixed\\
\hline 
\mhu        	& GUT-scale soft mass of $\hu$          & 0.1, 4 (0.1, 20)	& Linear\\
\mhd		& GUT-scale soft mass of $\hd$          & 0.1, 4 (0.1, 20) & Linear\\
\hline 
\multicolumn{4}{|l|}{Nuisance parameters like in the CMSSM} \\
\hline
\end{tabular}

\footnotesize{$\ast$ In parentheses we show the ranges for the scans giving the 1TH region, see Sec.~\ref{sec:nuhm}.}
\caption{Priors for the parameters of the NUHM  and for the SM nuisance
  parameters used in our scans. Soft masses and \azero\ are in
  TeV. Top quark pole mass and bottom quark mass are in GeV.}
\label{tab:priorsNUHM}
\end{centering}
\end{table}
\section{Results}\label{Present}

In this section we will present our numerical results. We will first
examine the impact of the current and the projected determination of
\brbsmumu\ on the different high probability regions of the parameter
space of the CMSSM and will discuss ensuing implications for testing
them. In particular, we will show that the AF region is likely to be
basically fully excluded if the SM (or else current) value of
\brbsmumu\ is confirmed with high precision. Next we will demonstrate
that future one-tonne detectors of dark matter scattering off nuclei
will provide a crucial complementary way of cross-examining those
regions and of potentially exploring the favored regions of the CMSSM
over very wide ranges of parameters not accessible to direct LHC
searches for new particles. Next we will apply a similar approach to
the NUHM and show that the above conclusions in general will not
hold. On the other hand, some positive measurements of DM signal will
have the potential to basically rule out the CMSSM.

\subsection{The CMSSM}\label{sec:cmssm}

\begin{figure}[t]
\centering
\subfloat[]{%
\label{fig:-a}%
\includegraphics[width=0.45\textwidth]{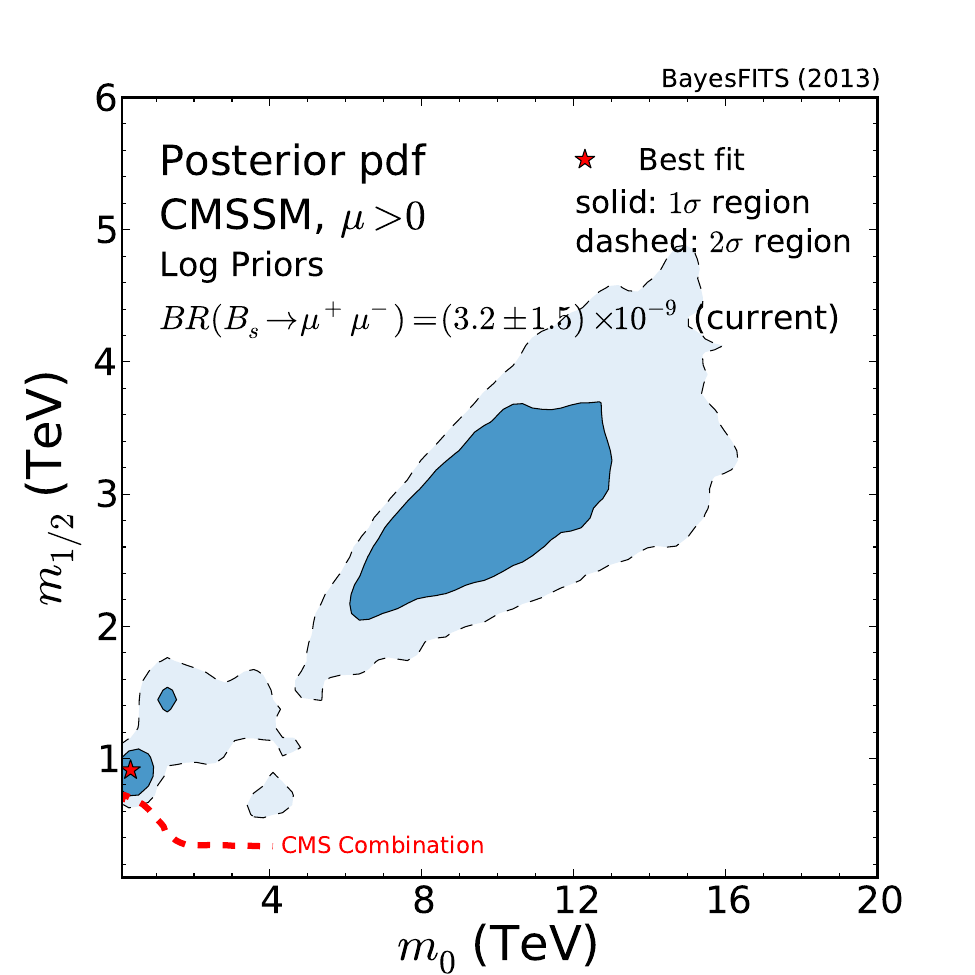}
}%
\subfloat[]{%
\label{fig:-b}%
\includegraphics[width=0.45\textwidth]{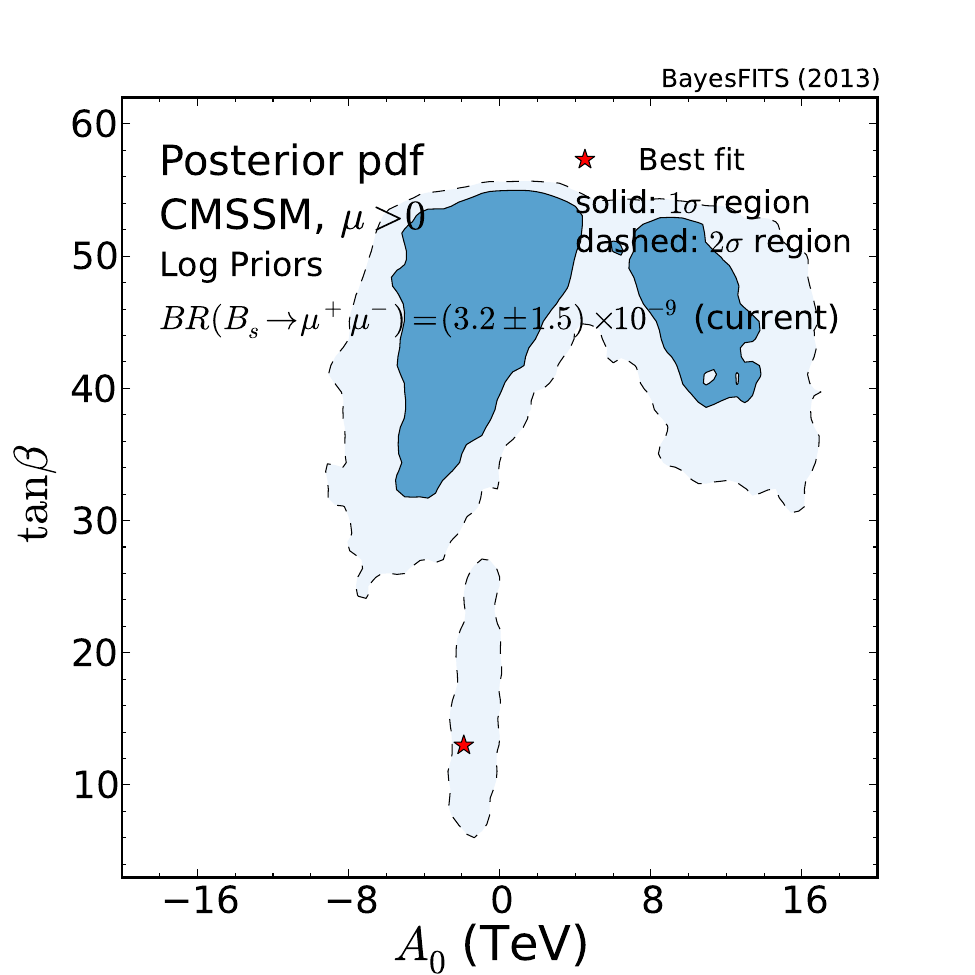}
}%
\hspace{1pt}%
\subfloat[]{%
\label{fig:-c}%
\includegraphics[width=0.45\textwidth]{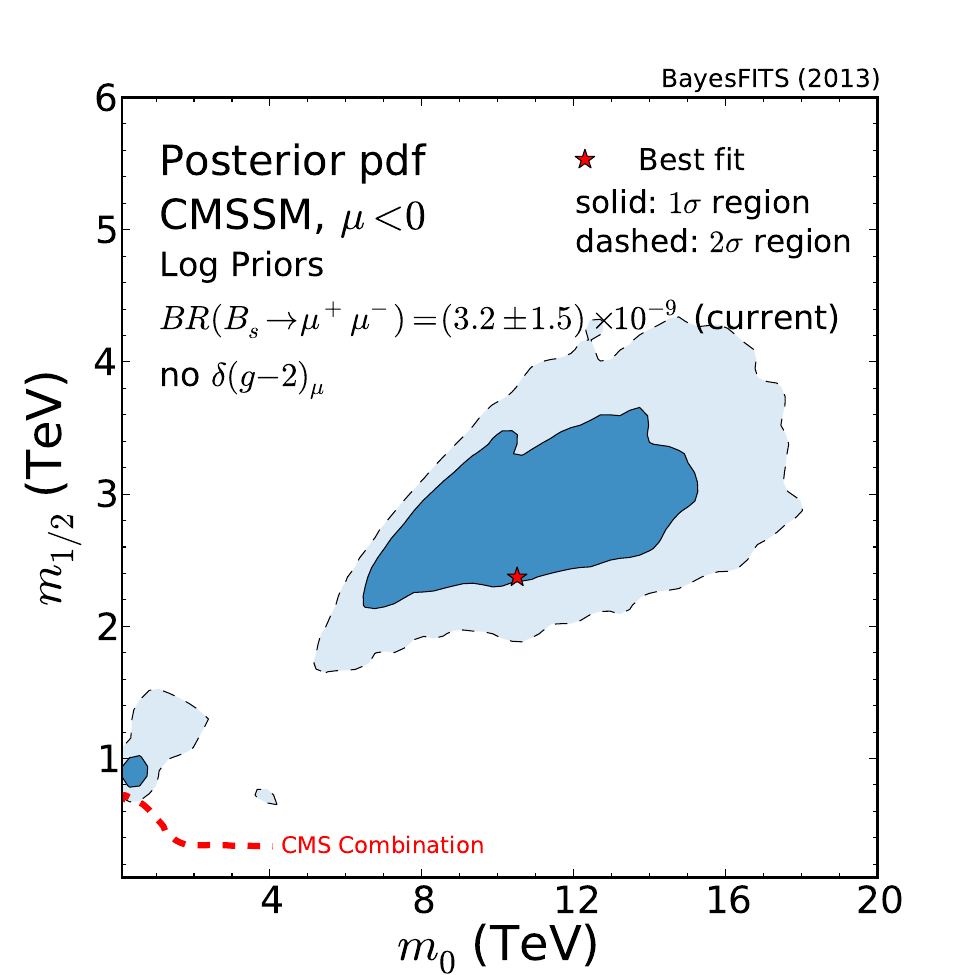}
}%
\subfloat[]{%
\label{fig:-d}%
\includegraphics[width=0.45\textwidth]{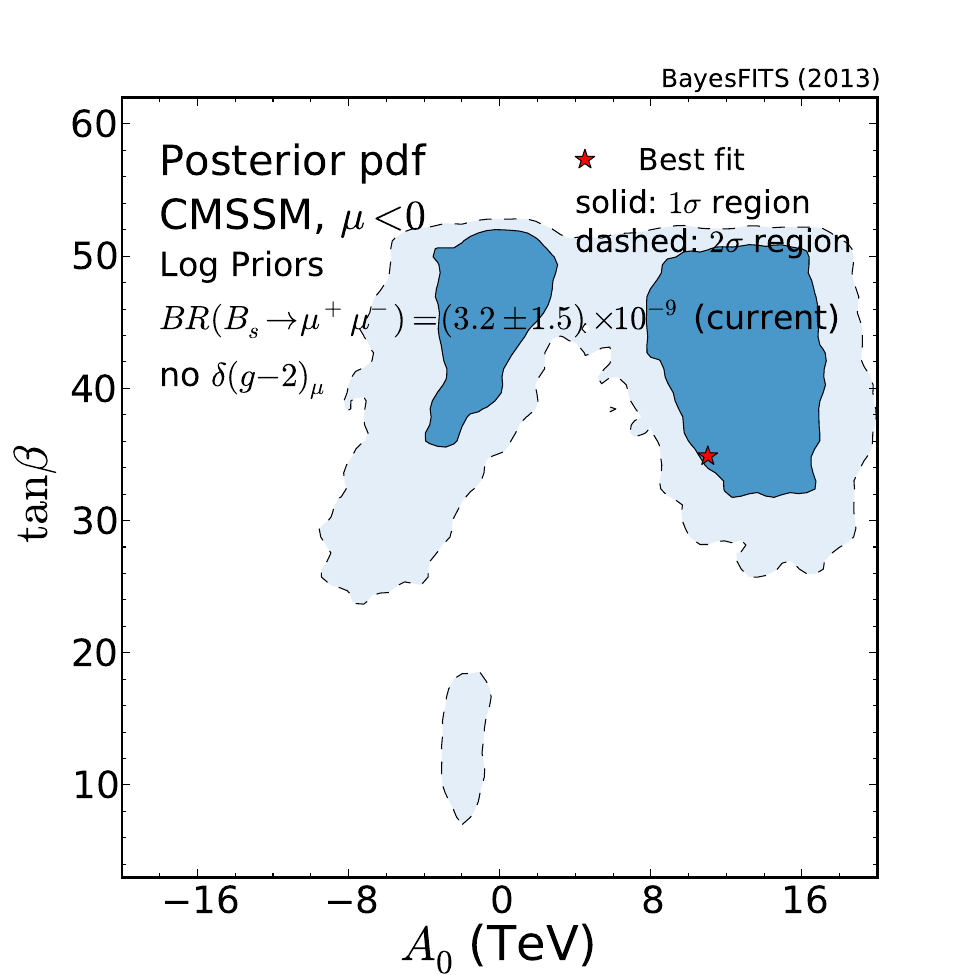}
}%
\caption[]{Marginalized 2D posterior pdf in \subref{fig:-a} the
  (\mzero, \mhalf) plane of the CMSSM for $\mu>0$, \subref{fig:-b}
  the (\azero, \tanb) plane for $\mu>0$, \subref{fig:-c} the (\mzero,
  \mhalf) plane for $\mu<0$, and \subref{fig:-d} the (\azero, \tanb)
  plane for $\mu<0$, constrained by the experiments listed in
  Table~\ref{tab:exp_constraints}, with the exclusion of
  \deltagmtwomu\ for $\mu<0$.  The 68\% credible regions are shown in
  dark blue, and the 95\% credible regions in light blue. The dashed
  red line shows the CMS combined 95\%~CL exclusion bound.}
\label{fig:cmssm_param}
\end{figure} 

In Figs.~\ref{fig:cmssm_param}\subref{fig:-a}
and~\ref{fig:cmssm_param}\subref{fig:-b} we plot 68\% and 95\%
credibility regions of a two-dimensional (2D) marginalized posterior
pdf (henceforth called posterior for brevity) in the (\mzero, \mhalf)
plane and in the (\azero, \tanb) plane, respectively, for $\mu>0$.  In
Figs.~\ref{fig:cmssm_param}\subref{fig:-c}
and~\ref{fig:cmssm_param}\subref{fig:-d} we show the same for $\mu<0$,
but without \deltagmtwomu, as mentioned earlier.  The figures give an
update and a significant extension of the results presented in our
previous CMSSM analysis\cite{Fowlie:2012im} by an inclusion of the new
positive measurement of \brbsmumu\ (instead of an upper limit) and by
significantly extending mass parameter ranges; compare
Table~\ref{tab:priorsCMSSM}.  In the ranges overlapping with those in
the previous study ($0.1\tev\leq \mzero\leq 4\tev$,
$0.1\tev\leq\mhalf\leq 2\tev$, $-7\tev\leq\azero\leq7\tev$ and
$3\leq\tanb\leq 62$) the figures basically reproduce the same
features, with the main three regions of high posterior favored
primarily by the DM relic density and the Higgs mass, and also by the
other constraints.  First, the SC region shows up at small \mzero\
just above the LHC (CMS, and similarly for ATLAS) exclusion
line.\footnote{The SC strip can be narrowed down by applying limits on
  long-lived charged particles to staus\cite{Citron:2012fg} but this
  will not significantly change the results presented here.}  The
posterior features a 68\% credibility and the best-fit point is
located there thanks to a very good fit to the Higgs mass, and a value
of \brbsmumu\ in agreement with the experiment (at not too large
\tanb). Next, the AF region can be seen at the $2\sigma$ credibility
level for $1\tev\lsim\mzero\lsim 4\tev$ and $1.2\tev\lsim\mhalf\lsim
2\tev$, although a much smaller $1\sigma$ `island' at smaller
\mzero\ is also present.  Finally, the FP/HB region appears only as a
95\% credibility island at $\mzero\approx 4\tev$ due to the fact
that it is more difficult there to produce the correct Higgs mass. (See\cite{Fowlie:2012im} for a
detailed discussion, and also\cite{Kowalska:2012gs} where we discussed
in detail the CMSSM limit of the CNMSSM, and adopted the same updated
values of experimental constraints as in this study.)
 
As a side remark, we note that in\cite{Fowlie:2012im} the best-fit
point was located in the AF region.\footnote{It was also emphasized
  there that the location of the best-fit point in the CMSSM is very
  sensitive to exact values of input parameters, approximations used,
  \etc.}  With the new improved fit the best-fit point is now found in
the SC region -- this is due to the updated (somewhat increased) value
of the top pole mass which made it easier to obtain a 126\gev\ Higgs
mass in the SC region, also in the CNMSSM, as we discussed in detail
in\cite{Kowalska:2012gs}.

In the case of $\mu<0$ (but without \deltagmtwomu) the AF region is
much less prominent than for $\mu>0$, although still visible in
Fig.~\ref{fig:cmssm_param}\subref{fig:-c} at 95\% credibility.
Likewise the FP/HB region has shrunk considerably, while the SC
remained fairly stable. 

Going to larger \mzero\ and \mhalf, beyond those considered
in\cite{Fowlie:2012im}, the main new feature in
Figs.~\ref{fig:cmssm_param}\subref{fig:-a} and
\ref{fig:cmssm_param}\subref{fig:-c} is the appearance of a large 68\%
posterior region ranging from around 5 to 12\tev\ in
\mzero\cite{Akula:2012kk} where the LSP is an almost purely
higgsino-like neutralino with mass $\mchi\approx\mu\simeq1\tev$ (the
1TH region). The correct Higgs mass is also easily obtained there due
to large \msusy\ while all other constraints, including \brbsmumu\
(except at large \tanb), reproduce basically the SM value there.  In
fact, for $\mu<0$ with \deltagmtwomu\ dropped from the list of
constraints, the best-fit point has now moved up to the 1TH region
since no other constraint favors lower \msusy.

Notice that Figs.~\ref{fig:cmssm_param}\subref{fig:-b}
and~\ref{fig:cmssm_param}\subref{fig:-d} show that the parameters
\azero\ and \tanb\ are now less constrained than
in\cite{Fowlie:2012im}.  This is a consequence of extending
the scanned ranges of \mzero\ and \mhalf\ to much larger values. The large higgsino DM
region corresponds to two large 68\% credible regions, where \tanb\
assumes values in the range 30--55, and \azero\ can take very large
negative and positive values. The tree level value of 
$\mhu^2$ is often positive, in which case electro-weak symmetry breaking (at the tree level) 
is not achieved. To overcome this, large and negative one-loop contributions to $\mhu^2$, which are proportional to 
$|A_t|\tanb$, are needed. Therefore, at smaller \tanb\ larger $|A_t|$ (and hence $|\azero|$) are favored -- a tendency
that becomes weaker as \tanb\ grows.
Hence, the posterior features a `gap' for small \azero, which narrows down with increasing
\tanb\ ($0\lesssim\azero\lesssim5\tev$ implies $|A_t|\ll\azero$ for most choices of the other parameters). 
Values of $\tanb\lesssim 25$, on the other hand, are not favored in this region given 
the prior ranges considered in this study. 

\begin{figure}[t!]
\centering
\subfloat[]{%
\label{fig:-a}%
\includegraphics[width=0.45\textwidth]{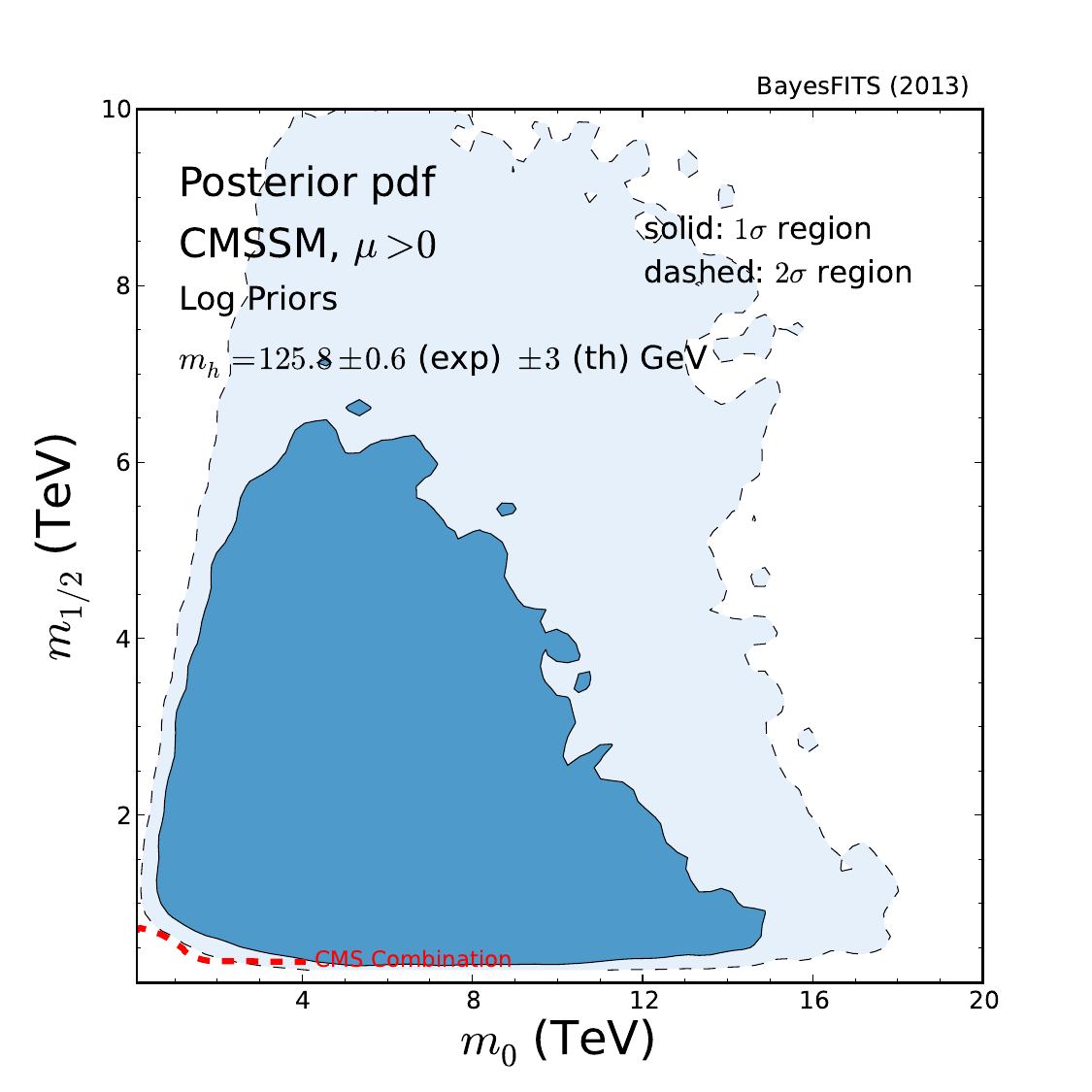}
}%
\subfloat[]{%
\label{fig:-b}%
\includegraphics[width=0.45\textwidth]{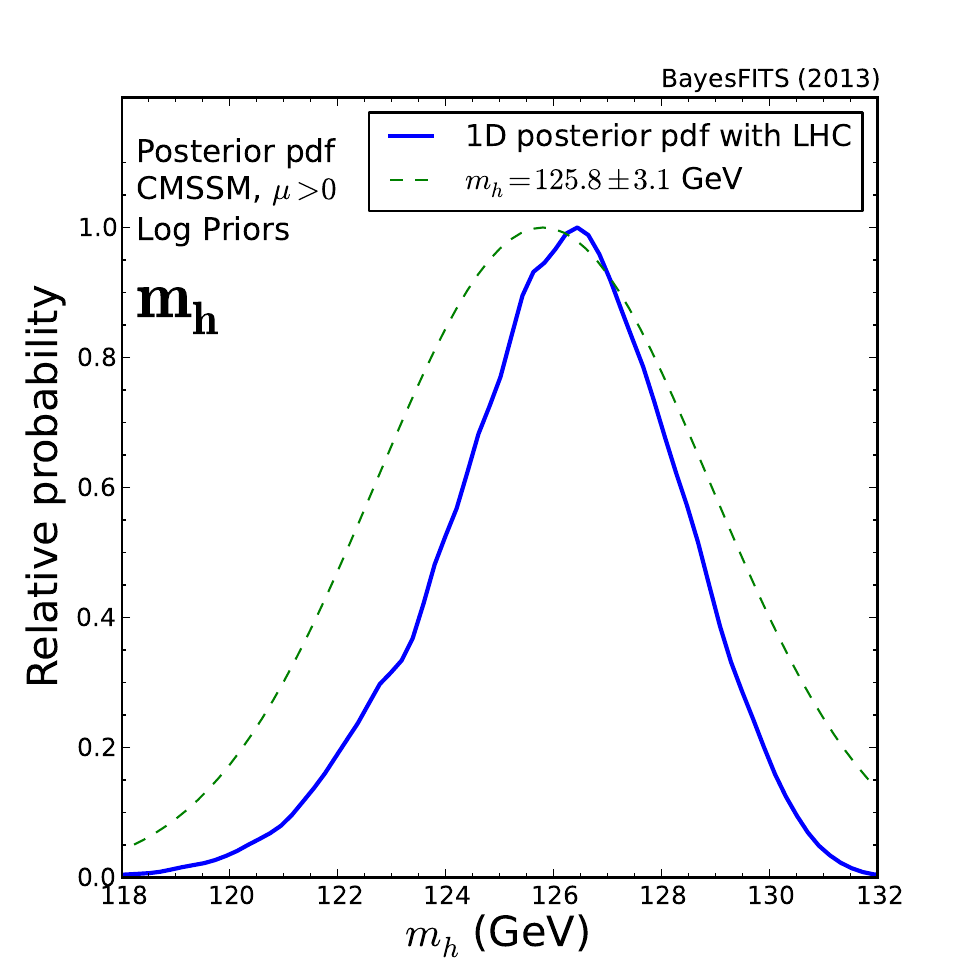}
}%
\caption[]{\subref{fig:-a} Marginalized 2D posterior pdf in the (\mzero, \mhalf) plane
  of the CMSSM constrained only by the Higgs mass and the LHC SUSY searches. The
  68\% credible regions are shown in dark blue, and the 95\% credible
  regions in light blue. The dashed red line shows the CMS combined
  95\%~CL exclusion bound. \subref{fig:-b} Marginalized 1D posterior pdf 
for \mhl\ (solid blue line) under the same assumptions as in \subref{fig:-a}. The dashed green line 
shows the Higgs mass likelihood, with experimental and theoretical uncertainties added in quarature.}
\label{fig:cmssm_higgs} 
\end{figure} 

At this point one can raise the question whether 
the 95\% credibility upper bound to the 1TH region ($\mzero\approx16\tev$, $\mhalf\approx 5\tev$)
is due to the physical impact of the constraints considered here, or is rather a feature of our choice 
of priors and parameter ranges.
We will explain this in what follows. 

In the regime of such large masses, SUSY contributions to all electroweak and flavor observables
become very small, 
so that the only constraints that can affect the favored parameter space are the relic density and the Higgs mass. 
It has being long known that the loop corrections to the Higgs mass increase logarithmically 
with increasing \msusy\ so that, in principle, the measured value of the Higgs mass
can place an upper limit on \mzero, \mhalf. To exemplify this feature we show in Fig.~\ref{fig:cmssm_higgs}\subref{fig:-a} 
the marginalized posterior
pdf in the (\mzero, \mhalf) plane for the parameter ranges considered in this analysis, 
in the case where all other constraints with the exception of the Higgs mass and 
SUSY limits from the LHC are turn off.
One can see that the 68\% credibility region does not extend 
beyond $\mzero\simeq 14\tev$ and $\mhalf\simeq 6\tev$.
In Fig.~\ref{fig:cmssm_higgs}\subref{fig:-b} we show the marginalized 1D posterior 
pdf for the Higgs mass in this case (solid blue line), to confirm that \mhl\ can reproduce the 
experimental value very well with these parameter ranges. Moreover, the nearly Gaussian shape of the 1D pdf
implies basically no tension with the CMS lower limit in the (\mzero,
\mhalf) plane, as the 68\% credibility region favored by the Higgs
mass favors multi-TeV scale for both CMSSM parameters. In other
words, the Higgs mass of around 126\gev\ typically implies \msusy\ in
the range of a few to several TeV.

However, the upper bound on \mzero\ and \mhalf\ shown in Fig.~\ref{fig:cmssm_higgs}\subref{fig:-a}
does depend on the assumed parameter range and on the prior distribution. 
We have checked that, by extending the parameter space to $\mzero,\mhalf=50\tev$ and $\azero=\pm 50\tev$,
the 68\% and 95\% credibility bounds in the (\mzero, \mhalf) plane extend by approximately 50\%
in both directions when maintaining log priors, and by 50\% in \mzero\ and a factor of two in 
\mhalf\ when switching to flat priors.
Furthermore, it was recently shown in a detailed study\cite{Giudice:2011cg}
that for values of \tanb\ lower 
than the ones considered in this study ($1\leq\tanb<3$, disfavored by the relic density
constraint) there is vitually no bound on \msusy\
due to the Higgs mass, up to GUT scale.
 
On the other hand, the relic density does impose a much stronger bound on the favored parameter space. 
In the high-mass region, the tree-level $\mu$ parameter and the one-loop tadpole corrections to its
value can both significantly exceed the 1\tev\ scale. 
Since the relic density constraint in the 1TH region requires $\mu\sim1\tev$, as explained above, 
tree-level and one-loop contributions should cancel each other with very high accuracy, 
which requires precise tuning of the model parameters. 
This also affects the stability of the solutions provided by the spectrum generators (see\cite{Belanger:2005jk} for a detailed discussion). 
In fact, for $\mzero>20\tev$ it becomes very difficult to generate spectra with $\mu\sim1\tev$, 
and it becomes virtually impossible for $m_0>40\tev$.
This causes an upper bound on the high-probability higgsino regions shown in Figs.~\ref{fig:cmssm_param}\subref{fig:-a}
and \ref{fig:cmssm_param}\subref{fig:-c}.

Notice that, given the very strong constraints from the relic abundance, the upper bound on 
the 1TH region is basically range and prior independent. We checked this with a supplementary scan, 
with all constraints included, in which we extended the parameter ranges up to 50\tev\ 
for \mzero\ and up to 20\tev\ for \mhalf,
with log and linear priors. The upper bound to the 1TH region 
in the (\mzero, \mhalf) plane remained virtually unchanged.

\begin{figure}[t]
\centering
\subfloat[]{%
\label{fig:-a}%
\includegraphics[width=0.45\textwidth]{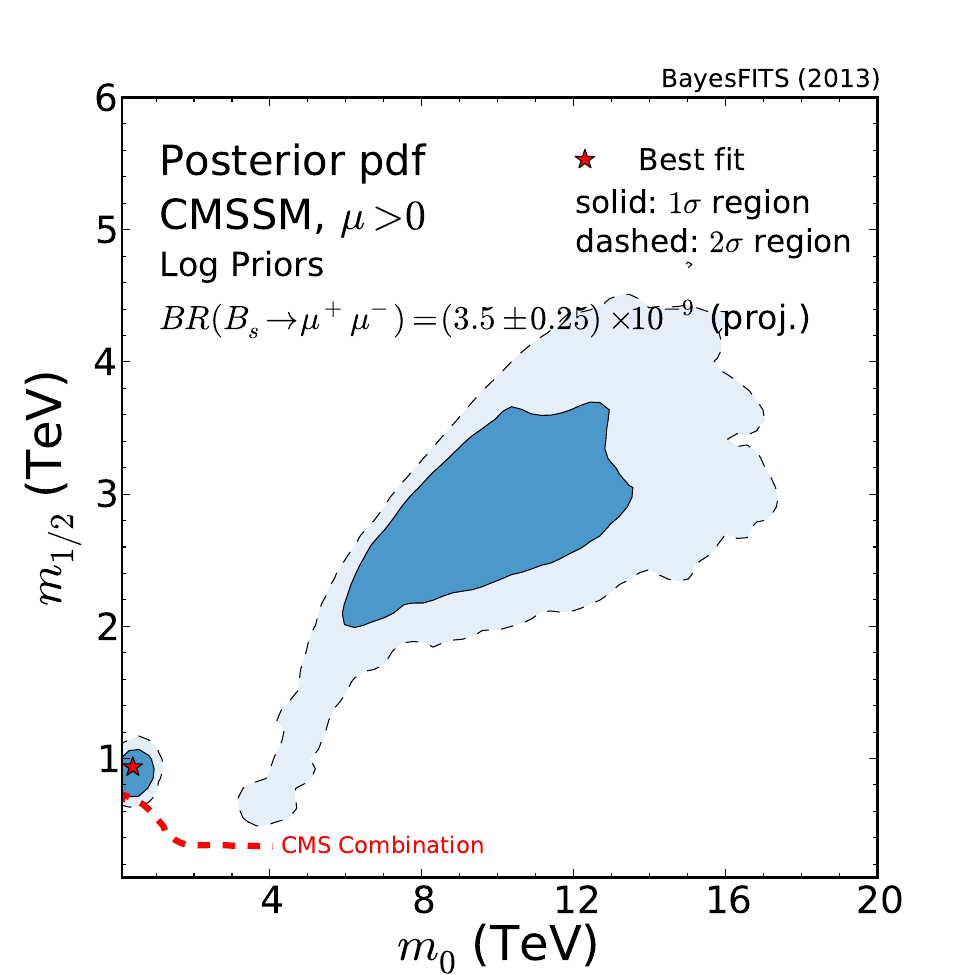}
}%
\subfloat[]{%
\label{fig:-c}%
\includegraphics[width=0.45\textwidth]{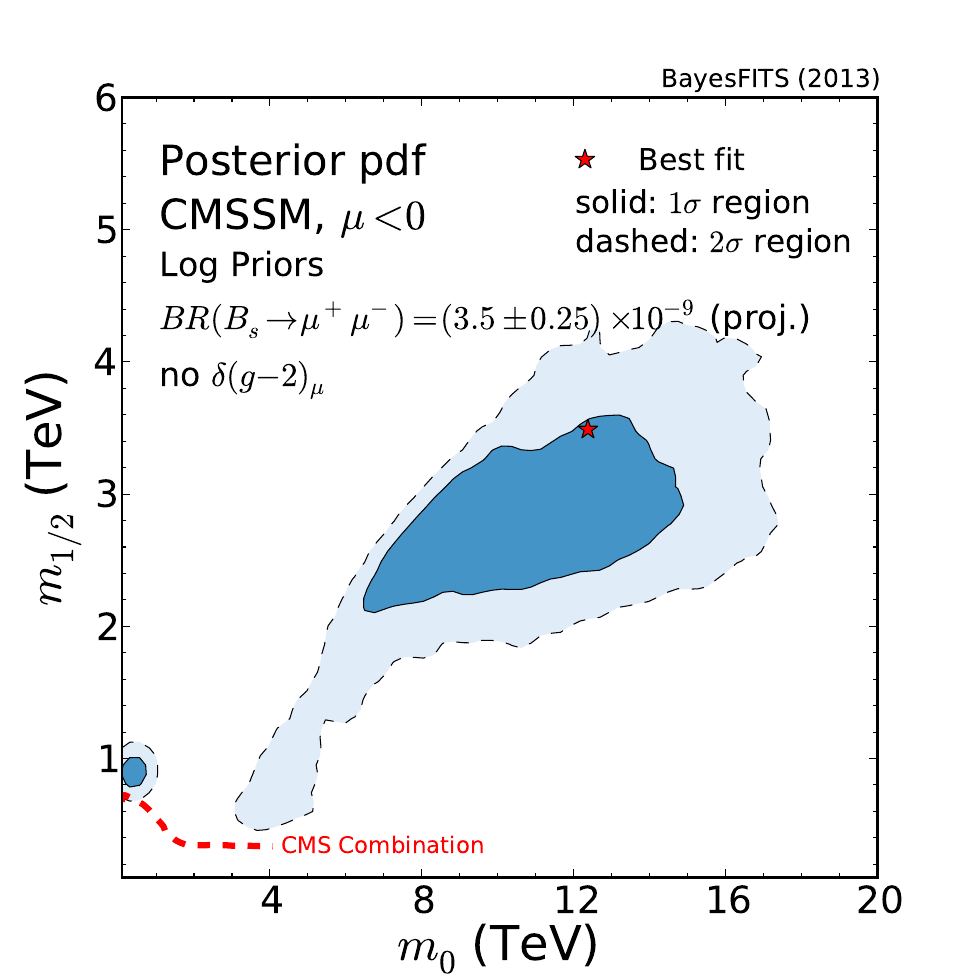}
}%
\caption[]{Marginalized 2D posterior pdf in the (\mzero, \mhalf) plane
  of the CMSSM constrained by the experiments listed in
  Table~\ref{tab:exp_constraints} with projected uncertainties for
  \brbsmumu. \subref{fig:-a} $\mu>0$, \subref{fig:-b} $\mu<0$. The
  68\% credible regions are shown in dark blue, and the 95\% credible
  regions in light blue. The dashed red line shows the CMS combined
  95\%~CL exclusion bound.}
\label{fig:cmssm_fut35}
\end{figure} 

In Figs.~\ref{fig:cmssm_fut35}\subref{fig:-a}
and~\ref{fig:cmssm_fut35}\subref{fig:-b} we show the marginalized 2D
posterior pdf in the (\mzero, \mhalf) plane for $\mu>0$ and $\mu<0$,
respectively, for the scans where we adopted projected future
theoretical and experimental uncertainties (added in quadrature)
for \brbsmumu, \ie, $\brbsmumu_{\textrm{proj}}=(3.5\pm 0.25)\times
10^{-9}$.  The AF region does not appear in the high posterior
anymore, even at 95\% credibility, while the other regions basically
do not change. This was to be expected in light of the discussion
presented in Sec.~\ref{Bsmu}.  The argument is valid for both signs of
$\mu$: when $\mu>0$ ($\mu<0$) \brbsmumu\ in the AF region assumes much
larger (smaller) values than the ones favored by the projected
uncertainties, as shown qualitatively in
Fig.~\ref{fig:bsmumu}\subref{fig:-d}.  The location of the best-fit
point is different for $\mu<0$ since, like in
Fig.~\ref{fig:cmssm_param}\subref{fig:-c}, the constraint from
\deltagmtwomu\ has not been included in the likelihood function.

The results shown in Fig.~\ref{fig:cmssm_fut35} are quite insensitive
to the projected uncertainties assumed for \brbsmumu. The shape of the
posterior pdf does not change even if they are doubled, which we have
checked numerically.  We also point out that the situation will not be
different if in the future the uncertainties are narrowed around the
currently measured central value ($3.2\times10^{-9}$) instead of the SM
value. We also checked this numerically, finding no significant difference in the resulting 
posterior.

\begin{figure}[t!]
\centering
\subfloat[]{%
\label{fig:-a}%
\includegraphics[width=0.45\textwidth]{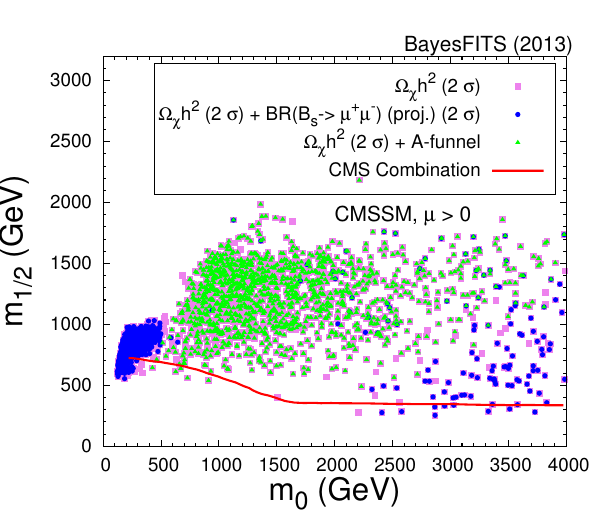}
}%
\subfloat[]{%
\label{fig:-b}%
\includegraphics[width=0.45\textwidth]{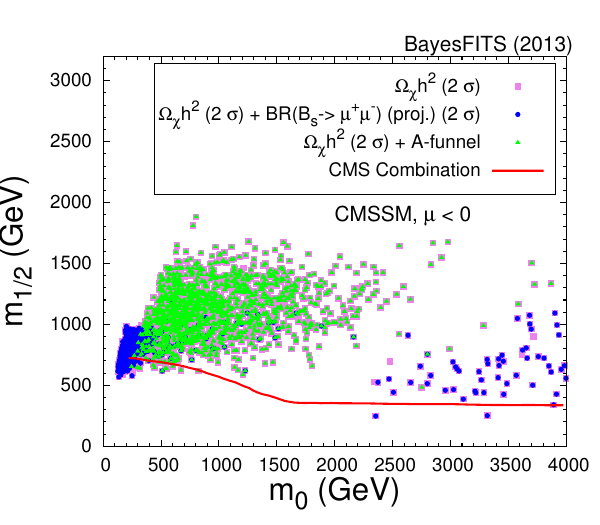}
}%
\caption[]{Scatter plot of the points in the (\mzero, \mhalf) plane of
  the CMSSM for \subref{fig:-a} $\mu>0$, and \subref{fig:-b}
  $\mu<0$, satisfying $\abundchi$ at 2$\sigma$ (pink squares),
  $\abundchi+\brbsmumu_{\textrm{proj}}$ at 2$\sigma$ (blue circles), and $\abundchi$
  at 2$\sigma$ and $|\ma-2\mchi|<100\gev$ (green triangles).}
\label{fig:cmssm_bsmu_afun}
\end{figure} 

To highlight the fact that in the CMSSM a more precise determination
of \brbsmumu\ can lead to an almost complete exclusion of the AF
region, we show in Fig.~\ref{fig:cmssm_bsmu_afun} scatter plots of the
points of our chains restricted to the low-mass regions. In
Fig.~\ref{fig:cmssm_bsmu_afun}\subref{fig:-a} $\mu>0$, whereas in
Fig.~\ref{fig:cmssm_bsmu_afun}\subref{fig:-b} $\mu<0$.  Pink squares
mark the points for which the relic density constraint is satisfied at
$2\sigma$ (theoretical + experimental errors added in quadrature); blue
circles represent the subset of these points for which \brbsmumu\ is
satisfied at projected $2\sigma$ (with total $\sigma=0.25\times
10^{-9}$); green triangles mark the subset of these points that belong
to the AF region ($|\ma-2\mchi|<100\gev$).  One can see a good spacial
separation between the blue and green points, which is a reflection of
the tension of the AF region with the \brbsmumu\ constraint.

The mass scales typical for the AF region are so high that most of it will
remain beyond direct reach of the LHC. (For example, with 300\invfb\ at
14\tev, CMS will probe $\mhalf\lsim1.3-1.4\tev$ at
$\mzero\lsim1\tev$\cite{Abdullin:1998pm}.)
Likewise, the FP/HB region will also be only partially probed at the
LHC, while the 1TH region will remain completely
beyond direct collider reach. As we have demonstrated, the projected
precision in the determination of \brbsmumu\ will have the power to
potentially rule out the AF region, but not the other ones.

\begin{figure}[t]
\centering
\subfloat[]{%
\label{fig:-a}%
\includegraphics[width=0.45\textwidth]{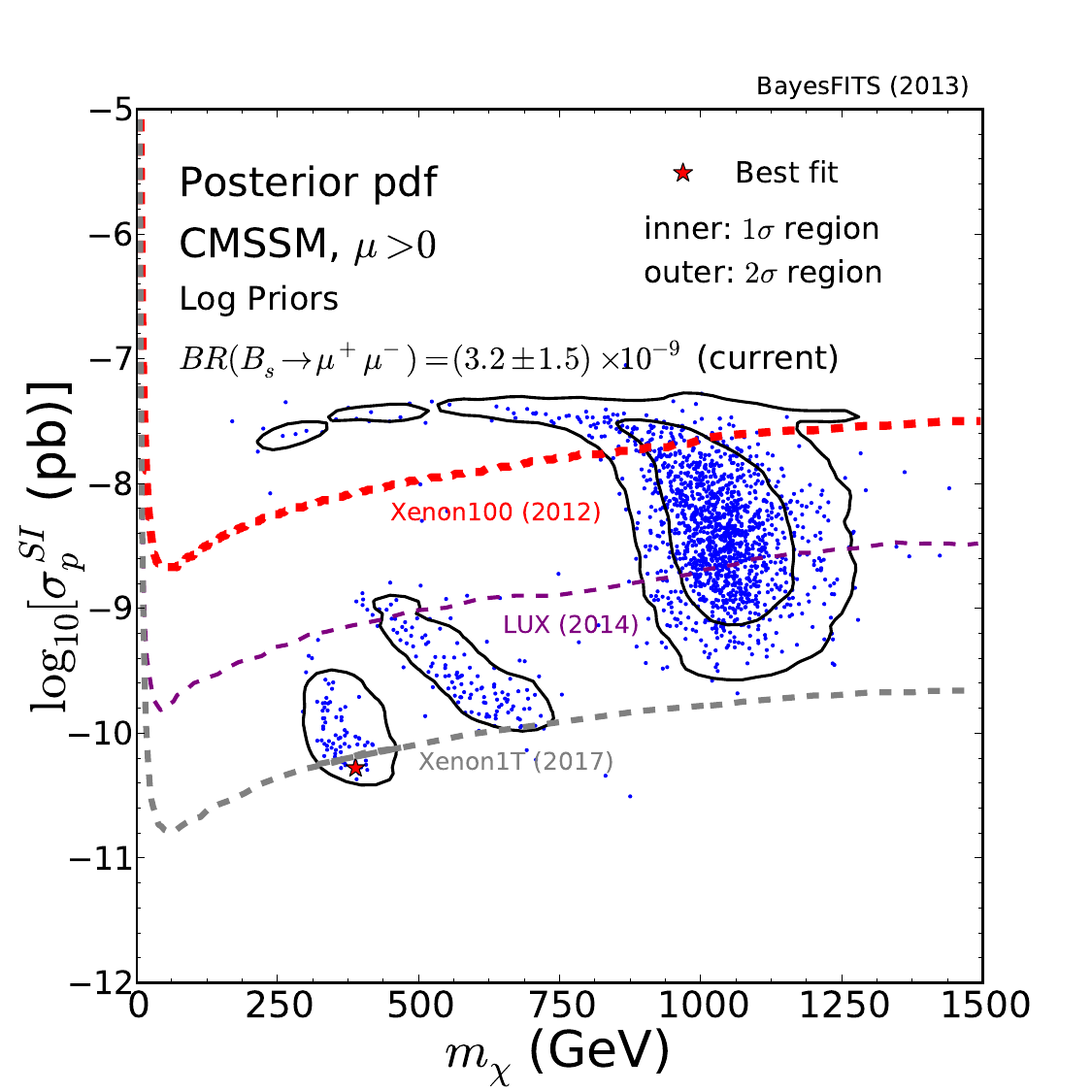}
}%
\subfloat[]{%
\label{fig:-b}%
\includegraphics[width=0.45\textwidth]{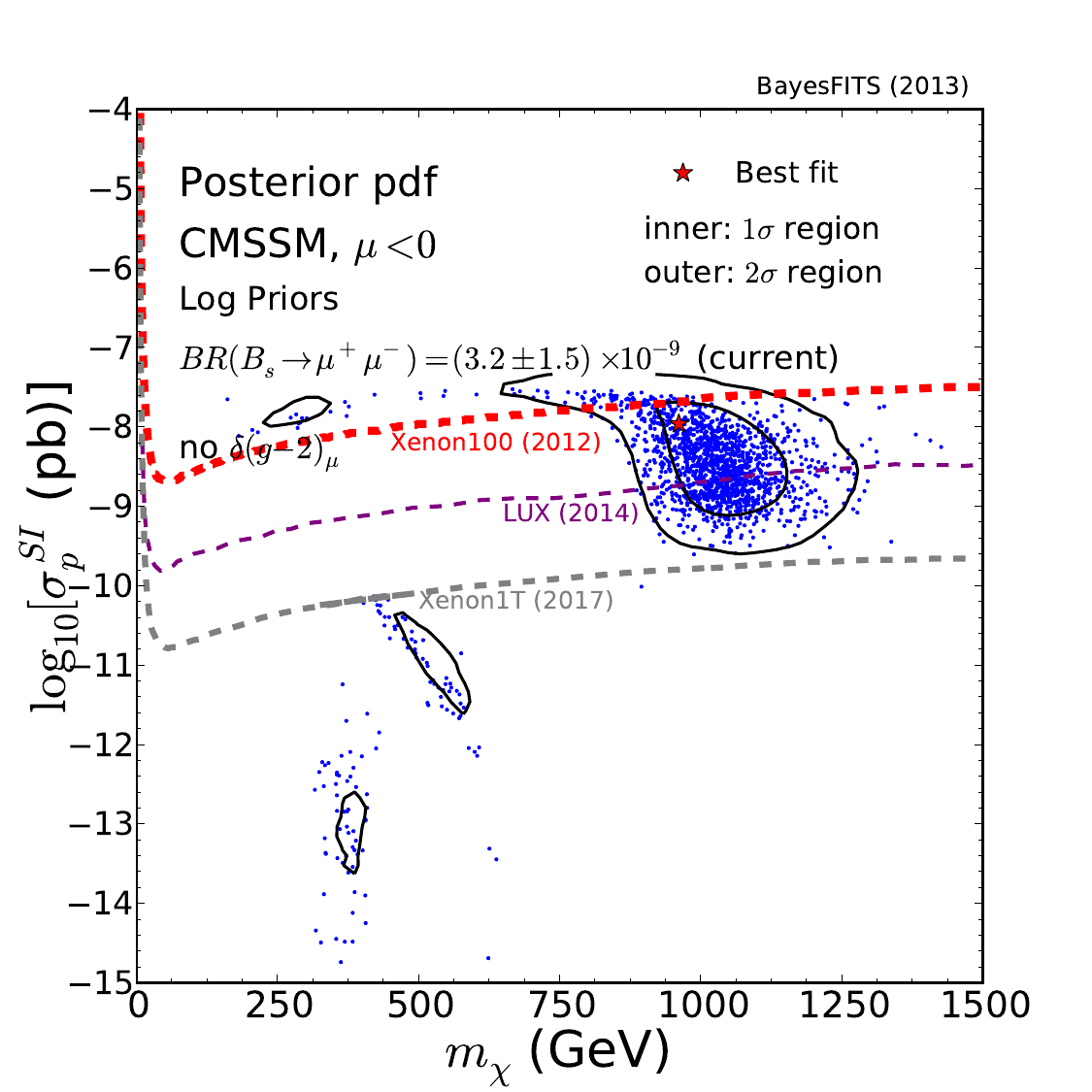}
}%
\hspace{1pt}%
\subfloat[]{%
\label{fig:-c}%
\includegraphics[width=0.45\textwidth]{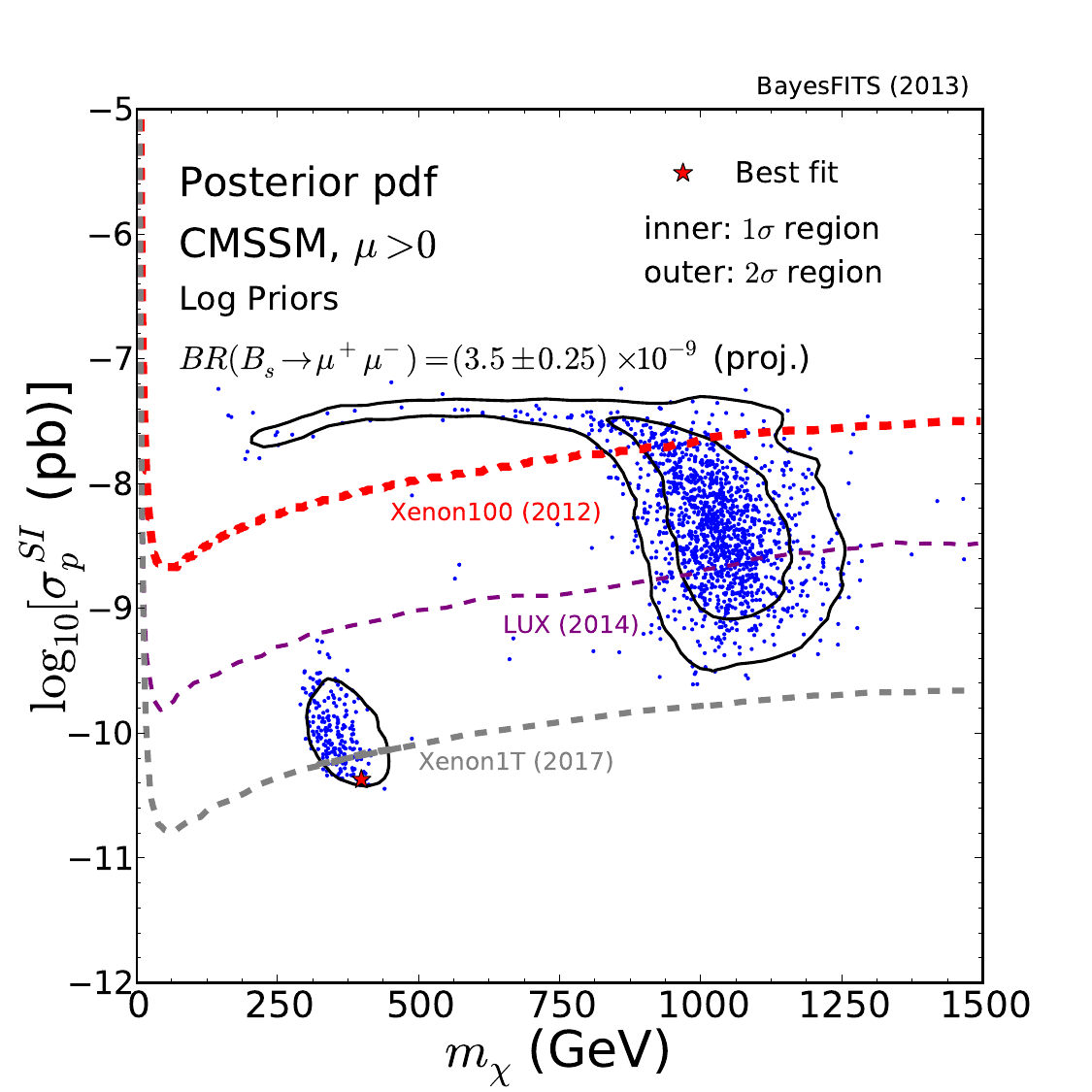}
}%
\subfloat[]{%
\label{fig:-d}%
\includegraphics[width=0.45\textwidth]{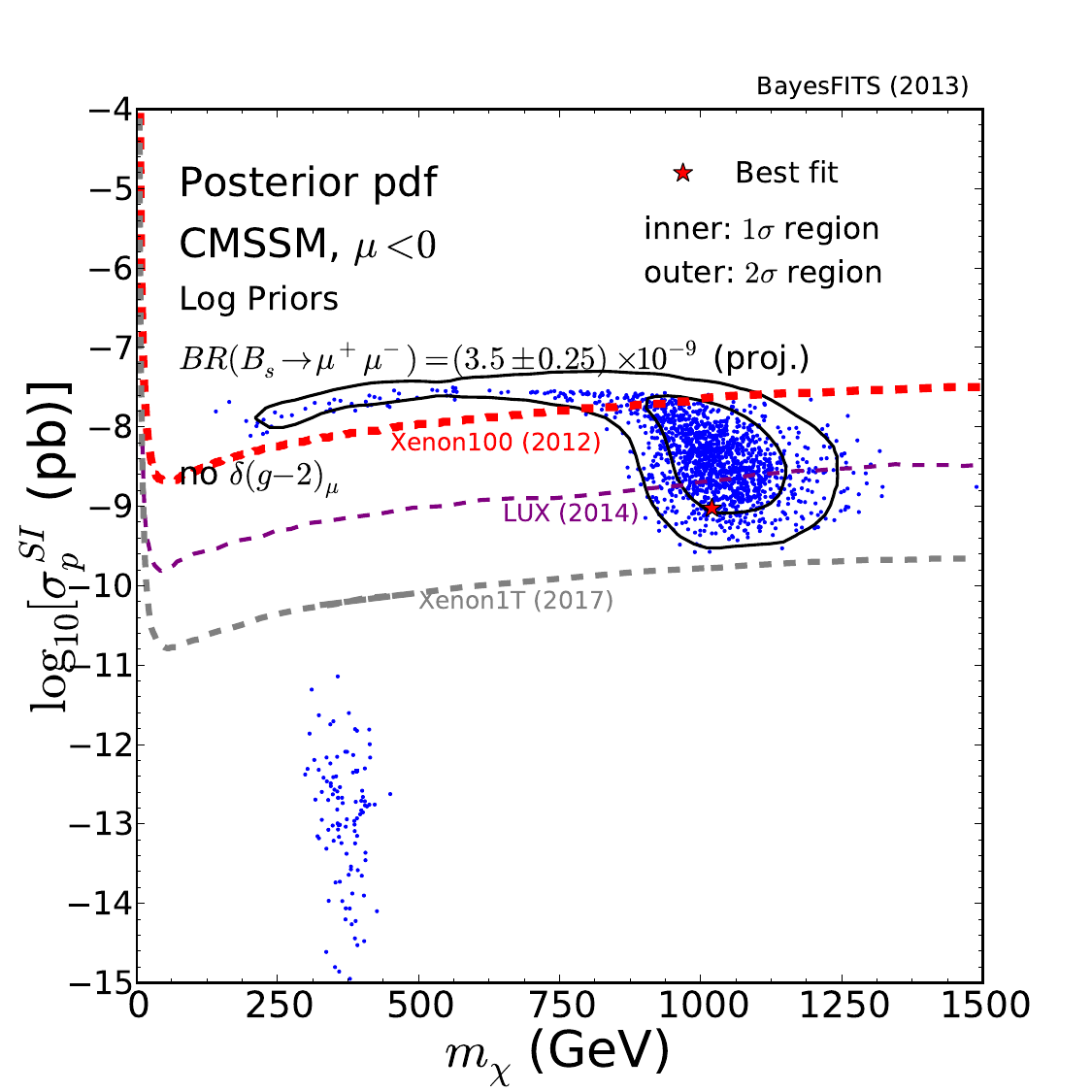}
}%
\caption[]{Marginalized 2D posterior pdf in the (\mchi, \sigsip) plane
  for the CMSSM constrained by the experiments listed in
  Table~\ref{tab:exp_constraints}. \subref{fig:-a} $\mu>0$, current
  uncertainties in \brbsmumu, \subref{fig:-b} $\mu<0$, current
  uncertainties in \brbsmumu, \subref{fig:-c} $\mu>0$, projected
  uncertainties in \brbsmumu, and \subref{fig:-d} $\mu<0$, projected
  uncertainties in \brbsmumu.  The dashed red line shows the 90\%~\cl\
  exclusion bound by XENON100 (not included in the likelihood), the dashed purple line the projected sensitivity
  for LUX, and
  the dashed gray line the projected sensitivity for XENON-1T.  A
  distribution of samples uniformly selected from our nested sampling
  chain is superimposed.}
\label{fig:cmssm_mchi_sig}
\end{figure} 

Fortunately, an expected ultimate sensitivity of DM searches in deep
underground detectors will provide a crucial complementary, and partly
overlapping, venue of testing all the high posterior probability
regions of the CMSSM.  In
Figs.~\ref{fig:cmssm_mchi_sig}\subref{fig:-a} and
\ref{fig:cmssm_mchi_sig}\subref{fig:-b} we show a 2D posterior in the
(\mchi, \sigsip) plane for $\mu>0$ and $\mu<0$, respectively.
Starting from Fig.~\ref{fig:cmssm_mchi_sig}\subref{fig:-a}, we can
clearly identify the four high posterior probability regions, each
with a characteristic LSP mass range and \sigsip. The SC region
(appearing only at 95\% credibility but featuring the best-fit point)
corresponds to fairly low \mchi\ ($\lsim450\gev$) and typically the
lowest \sigsip. Next to it, with somewhat larger \mchi\ and \sigsip\
lies the AF region (also at $2\sigma$ credibility), which in turn is
very well separated from the big 68\% credibility region of
$\sim1\tev$ higgsino LSP. Finally, the spin-independent cross section
in the FP/HB region featuring a mixed bino-higgsino neutralino (a
horizontal branch at $\sigsip\simeq3\times 10^{-8}\pb$) already shows
tension with the current 90\% \cl\ upper bound from
XENON100\cite{Aprile:2012nq}.  However, as demonstrated
in\cite{Roszkowski:2012uf} and mentioned above, this region is
probably not yet firmly excluded due to large theoretical and
astrophysical uncertainties.  The remaining three regions are
currently below the XENON100 exclusion line but will be almost
entirely probed by future detectors, as the projected
sensitivity lines for LUX\cite{Akerib:2012ys} and XENON-1T\cite{Aprile:2012zx} indicate.  
Note that, in the absence in the likelihood function of any constraint to
favor the SC or AF regions, the broad ranges of the CMSSM input
parameter assumed for our scans make the posterior strongly favor the 1TH
region, which presents the vast majority of points (the volume effect)
even with the log prior on \mzero\ and \mhalf,
although at 95\% of total posterior probability, the other regions are
also present.  We also note that in random scans one can find points
with reasonably good \chisq\ ($\delta\chisq\leq12$) lying beyond
those favored regions. We illustrate this by superimposing on the
posterior a distribution of samples uniformly selected from our nested
sampling chain (blue dots).

\begin{figure}[t]
\centering
\subfloat[]{%
\label{fig:-a}%
\includegraphics[width=0.45\textwidth]{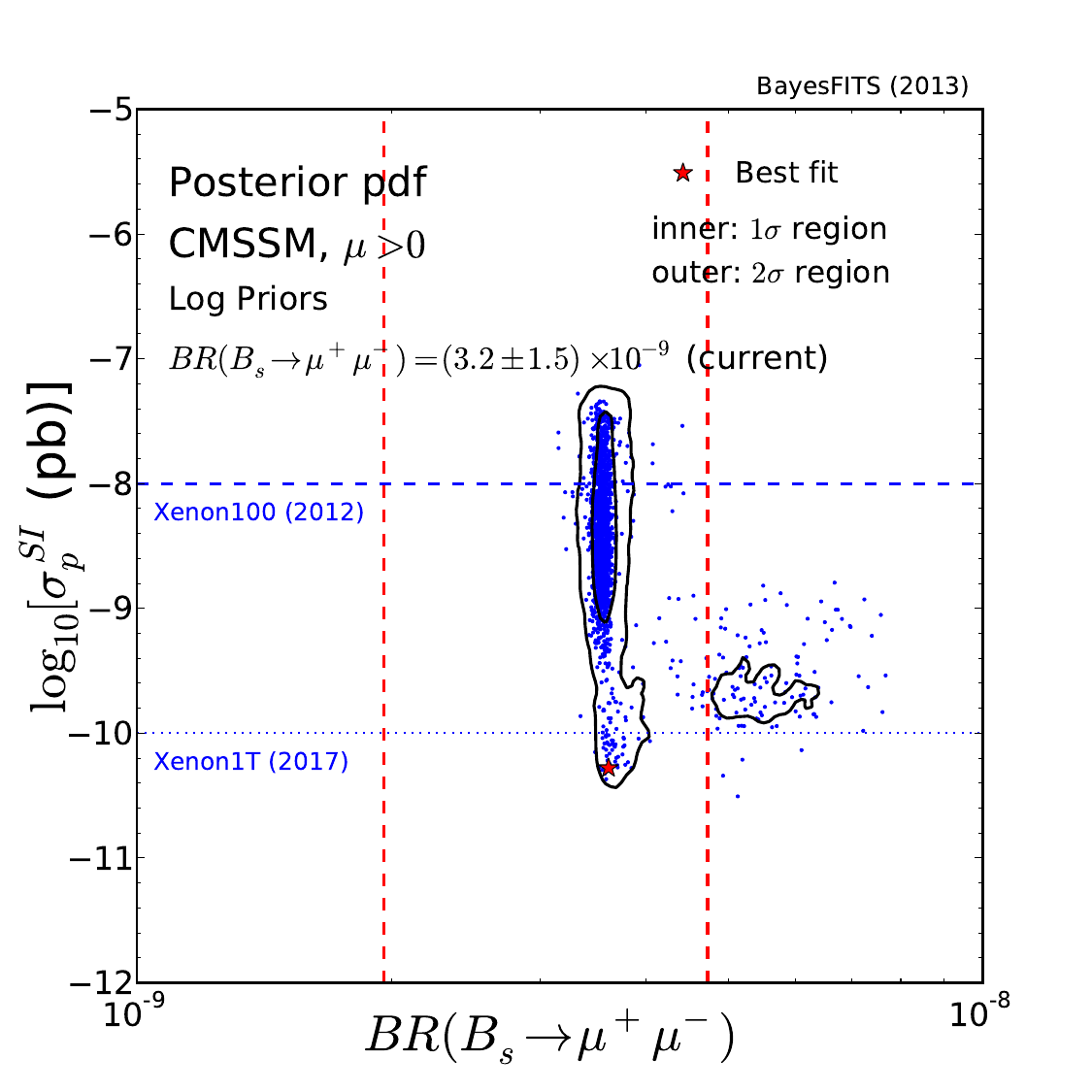}
}%
\subfloat[]{%
\label{fig:-b}%
\includegraphics[width=0.45\textwidth]{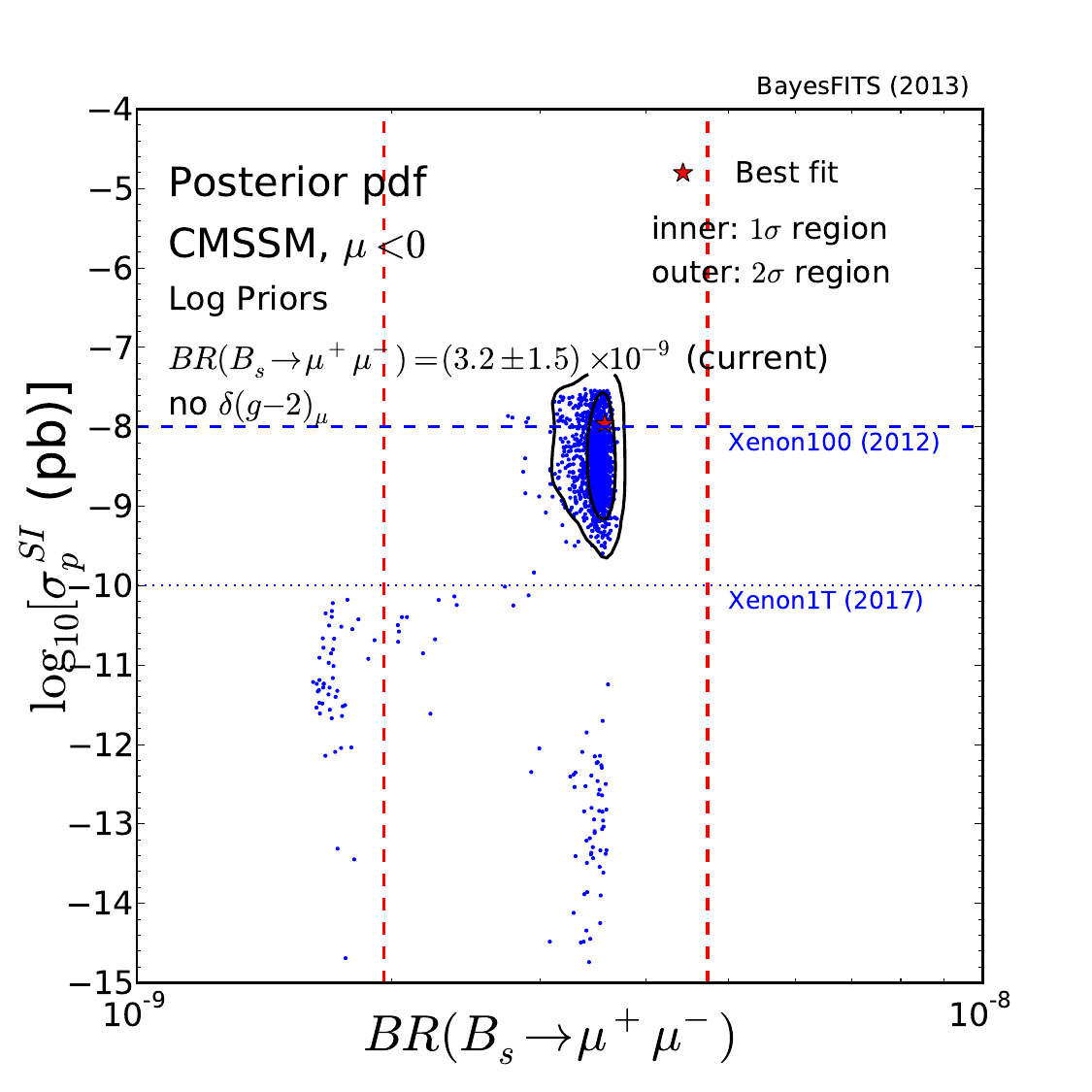}
}%
\hspace{1pt}%
\subfloat[]{%
\label{fig:-c}%
\includegraphics[width=0.45\textwidth]{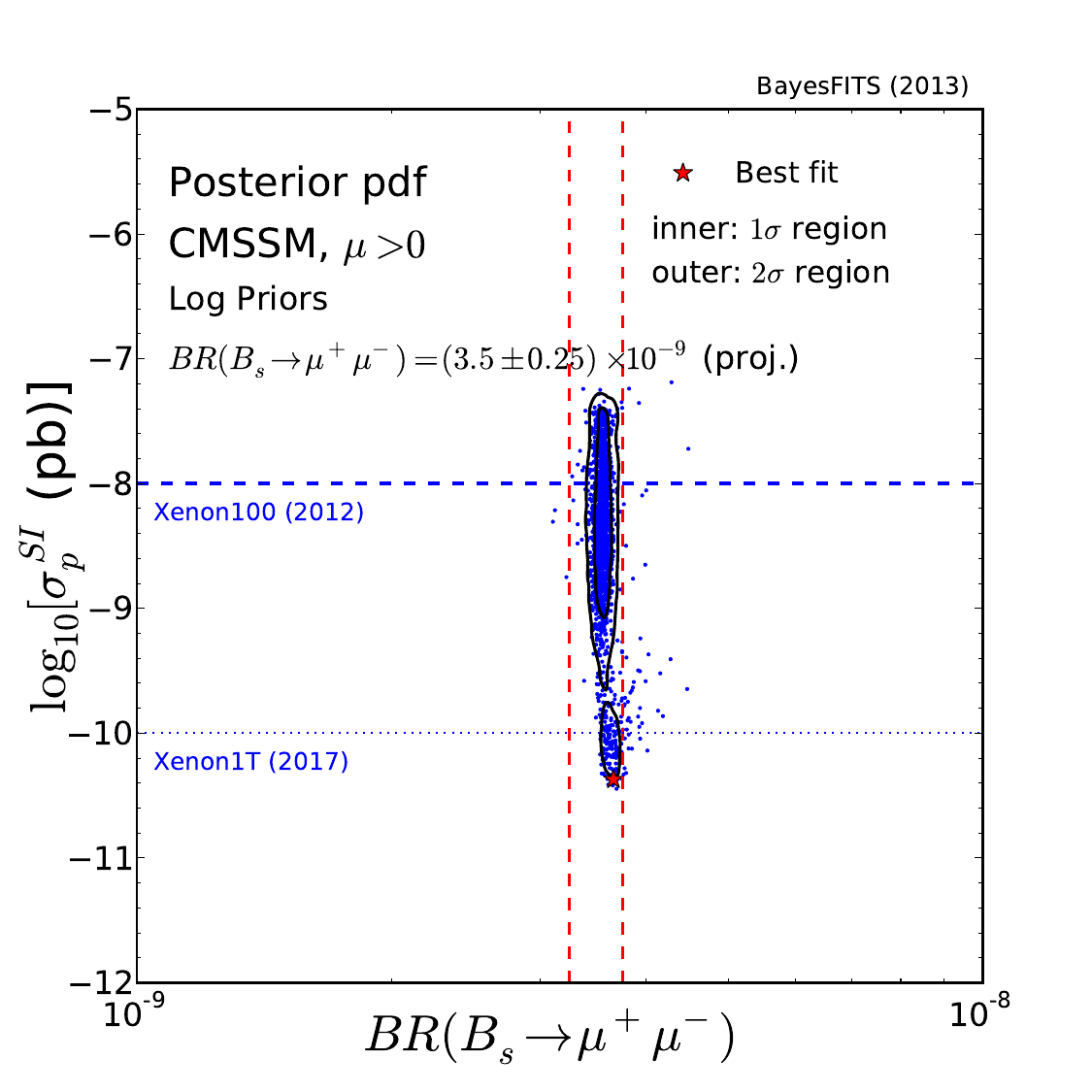}
}%
\subfloat[]{%
\label{fig:-d}%
\includegraphics[width=0.45\textwidth]{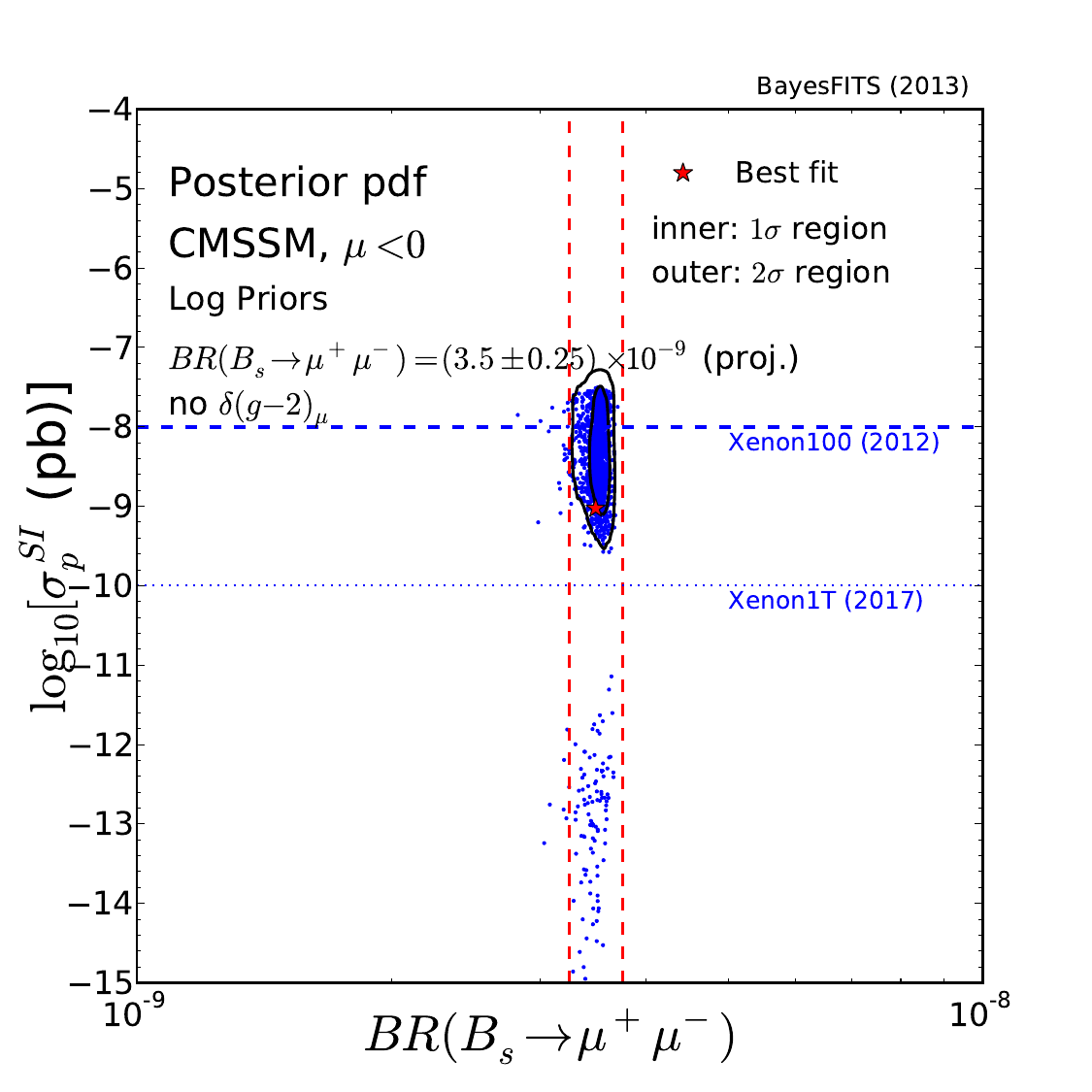}
}%

\caption[]{Marginalized 2D posterior pdf in the (\brbsmumu, \sigsip)
  plane for the CMSSM constrained by the experiments listed in
  Table~\ref{tab:exp_constraints}. \subref{fig:-a} $\mu>0$, current
  uncertainties in \brbsmumu, \subref{fig:-b} $\mu<0$, current
  uncertainties in \brbsmumu, \subref{fig:-c} $\mu>0$, projected
  uncertainties in \brbsmumu, and \subref{fig:-d} $\mu<0$, projected
  uncertainties in \brbsmumu.  The dashed red vertical lines show the
  current [\subref{fig:-a} and \subref{fig:-b}] and projected
  [\subref{fig:-c} and \subref{fig:-d}] uncertainties on \brbsmumu\ at
  $1\sigma$.  A distribution of samples uniformly selected from our
  nested sampling chain is superimposed. The dashed horizontal lines show the minimum 90\%~CL upper bound on 
  \sigsip\ by XENON100 (not included in the likelihood), and the dotted horizontal lines the corresponding
projected sensitivity for XENON1T.}
\label{fig:cmssm_bsmu_sig}
\end{figure} 

For comparison with the situation at present,
Figs.~\ref{fig:cmssm_mchi_sig}\subref{fig:-c} and
\ref{fig:cmssm_mchi_sig}\subref{fig:-d} show the same posterior in the
case where the future projected uncertainties on \brbsmumu\ are
assumed; in other words \brbsmumu\ is assumed to be basically
reproducing the SM value. For $\mu>0$ the AF region is now gone and
there remain essentially two testable regions: the 1TH
region, which should be basically fully reachable by future DM
searches only, and the SC region, testable also in part by direct
searches at the LHC.  Furthermore, they are so widely separated in the
plane that a detection of a DM signal, even with poor initial
determination of both \mchi\ and \sigsip, would have the power to
discriminate between them. Furthermore, for $\mu<0$ the CMSSM predicts
that only the higgsino region will be reachable by one-tonne
detectors, while in the SC region a well known cancellation of two
terms reduces \sigsip\ to hopelessly low values. This actually gives
one a chance, even if somewhat indirect one, to additionally determine
the sign of $\mu$ since any DM measurement indicative of the SC region
would most likely favor the positive sign of $\mu$.

Some of the points made above are recast in a somewhat different way in
Fig.~\ref{fig:cmssm_bsmu_sig} where we plot 2D posterior regions in
the $(\brbsmumu,\sigsip)$ plane assuming the current (upper panels)
and projected (lower panels) determination of \brbsmumu, as indicated
with vertical bars showing the combined (theory + experimental)
errors.  At present the AF region lies (for both signs of $\mu$)
mostly (at 95\% credibility level) beyond the current $1\sigma$
experimental lines, and clearly not yet firmly excluded.  However,
after the projected uncertainties on \brbsmumu\ are assumed, as shown
in Figs.~\ref{fig:cmssm_bsmu_sig}\subref{fig:-c} and
\ref{fig:cmssm_bsmu_sig}\subref{fig:-d}, only the two testable regions
mentioned above survive, the 1TH region corresponding to
larger \sigsip\ and the SC region at the borderline of XENON-1T reach
($\mu>0$) or below it ($\mu<0$).

\subsection{The NUHM}\label{sec:nuhm}

We have demonstrated above that projected but realistic sensitivities
of \brbsmumu\ will have the discriminating power to basically rule out
the AF region in the CMSSM. Furthermore, future one-tonne detectors of
DM will reach down to values of \sigsip\ such that either a signal in
one of the two remaining high posterior probability regions is
detected, or the CMSSM will basically be ruled out over very wide
ranges of its parameters (with the exception of the SC region at
negative $\mu$), thus reaching far above the direct sparticle mass
reach at the LHC. On the other hand, by detecting a DM signal at low
\mchi\ the sign of $\mu$ could potentially also be determined.

The question arises whether such rather strong statements extend
beyond the CMSSM. Unfortunately, it is easy to see that this is not
the case already in the NUHM, which is one of the simplest extensions
of the CMSSM. As mentioned above, in the NUHM, one can choose \mha\
and $\mu$ as the additional two free parameters; see
Eqs.~(\ref{ewsb1:eq})-(\ref{ewsb2:eq}). These are precisely the quantities
that played the crucial role in the CMSSM where they were, however,
tightly constrained. On the other hand, we will show that in the NUHM one predicts some
signatures for DM searches in one-tonne detectors that are absent in
the CMSSM - this could provide the way for ruling out the latter model
over multi-TeV ranges of mass parameters.

\begin{figure}[t]
\centering
\subfloat[]{%
\label{fig:-a}%
\includegraphics[width=0.45\textwidth]{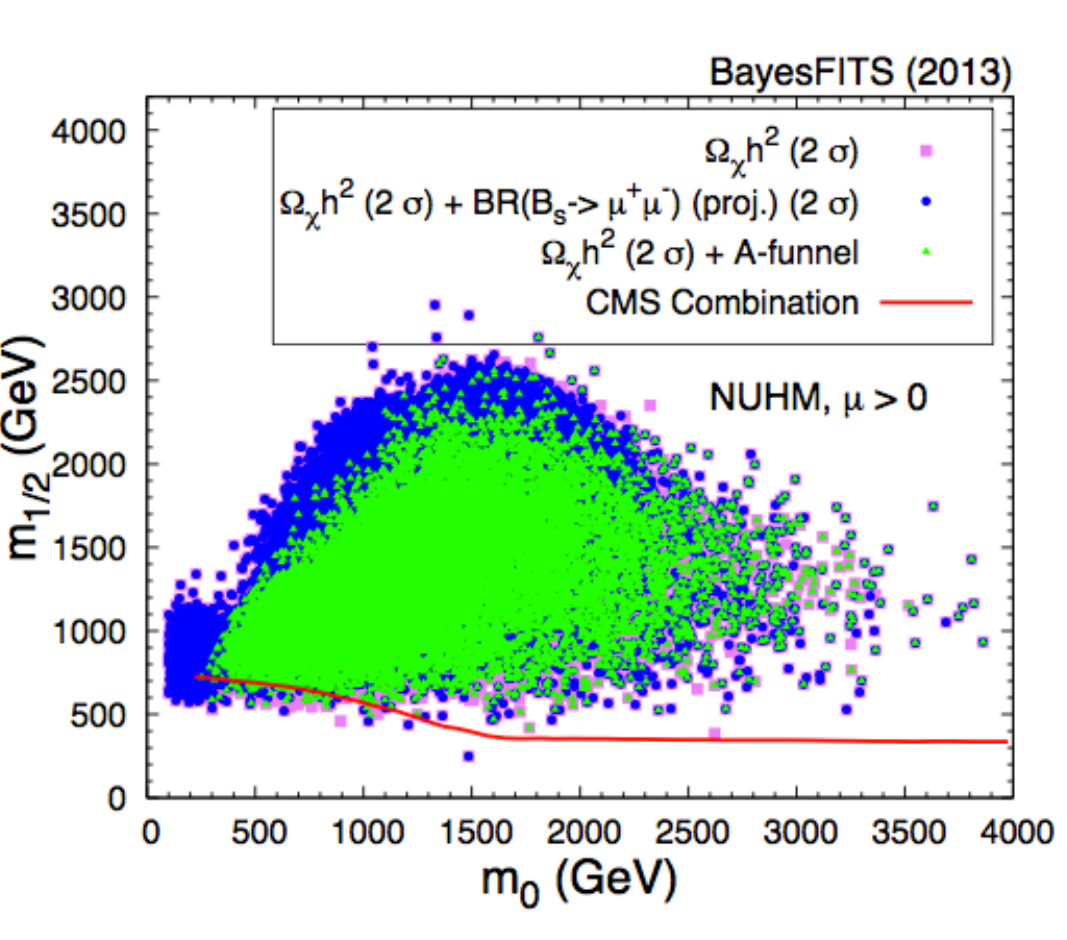}
}%
\subfloat[]{%
\label{fig:-b}%
\includegraphics[width=0.45\textwidth]{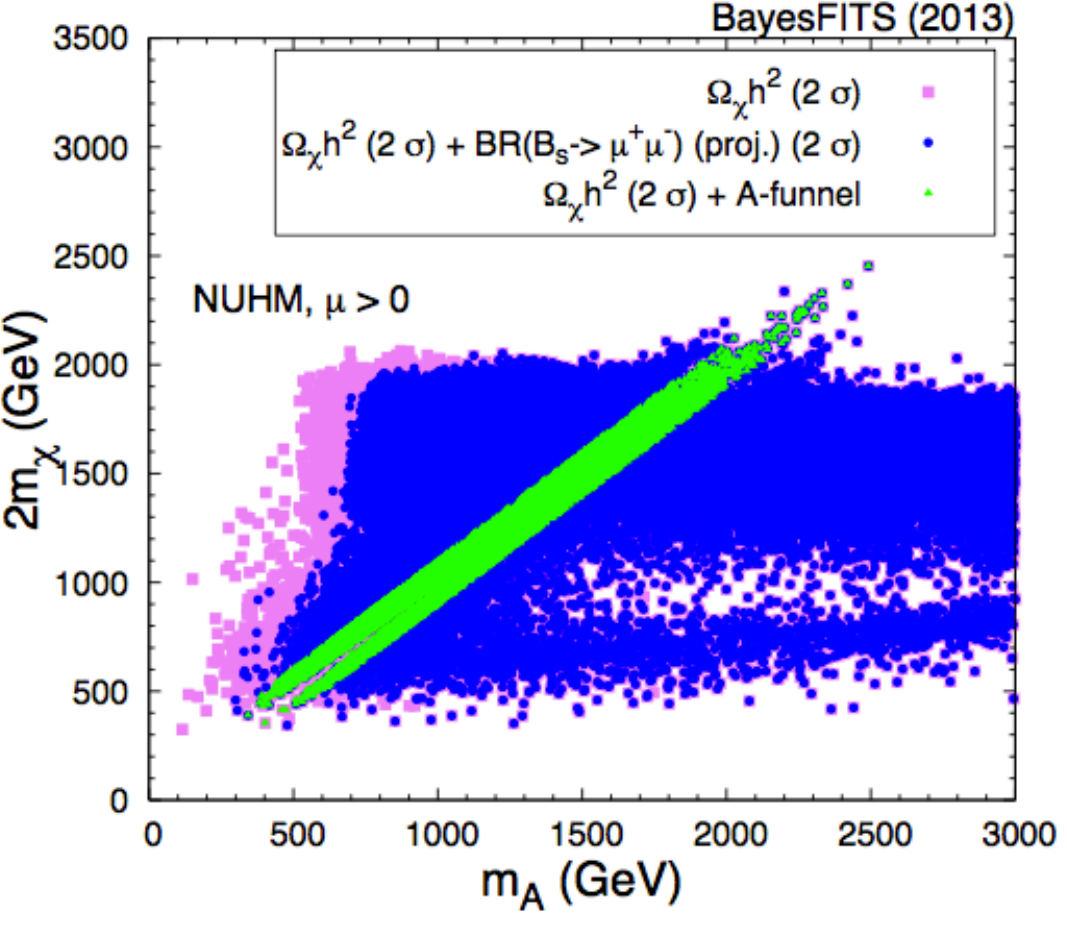}
}%
\caption[]{Scatter plot of the points in the \subref{fig:-a} (\mzero,
  \mhalf) plane and \subref{fig:-b} (\mha, $2\mchi$) plane of the NUHM
  for $\mu>0$ that satisfy $\abundchi$ at 2$\sigma$ (pink squares),
  $\abundchi +\brbsmumu_{\textrm{proj}}$ at 2$\sigma$ (blue circles), and $\abundchi$
  at 2$\sigma$ and $|\ma-2\mchi|<100\gev$ (green
  triangles).}
\label{fig:nuhm_bsmu_afun1}
\end{figure} 

Since the enlarged parameter space of the NUHM, with much more freedom
in the Higgs sector, allows a very good fit to almost all observables
(except invariably \deltagmtwomu), it is very time consuming to perform
a global Bayesian scan as above for the CMSSM. Additionally one has to worry
about much stronger prior dependence and volume
effect\cite{Roszkowski:2009sm}. However, since our goal in this paper
is to examine the impact of future \brbsmumu\ and direct DM search
sensitivities, a scan over a much more limited range of priors, given
in Table~\ref{tab:priorsNUHM}, is sufficient to provide a counter-example
to the conclusions drawn above in the CMSSM. Furthermore, we will not
need to draw Bayesian high posterior regions to make our point.

In Fig.~\ref{fig:nuhm_bsmu_afun1}\subref{fig:-a} we present, for
$\mu>0$, the distribution of points (pink squares) in the (\mzero,
\mhalf) plane for which the value of the relic density does not exceed the
central value by more than $2\sigma$.  (Since the relic abundance is a
strong constraint with a very small uncertainty, the distribution of
points determines 95\% credibility regions of the 2D pdf to very good
accuracy but we don't show them here.)  In green we show the subset of
these points for which the correct relic density is obtained through
neutralino annihilation via the $A$-resonance. These points constitute
the AF region of the NUHM.  We also show in blue the subset of the
pink points that will additionally satisfy the constraint on \brbsmumu\
within the projected $2\sigma$ error. In
Fig.~\ref{fig:nuhm_bsmu_afun1}\subref{fig:-b} we show the same sets of
points in the (\mha, $2\mchi$) plane, to highlight the features of the
AF region.  A very similar pattern emerges for $\mu<0$, hence we do
not show it here.

By comparing these figures with Fig.~\ref{fig:cmssm_bsmu_afun}, one
can see that, in contrast to the CMSSM, in the NUHM a precise
determination of \brbsmumu\ will have no
real discriminating power over the regions of the (\mzero, \mhalf)
plane, as the points do not show spacial separation. In other words,
the AF region will remain prominently allowed even if a future
determination of \brbsmumu\ will narrow it down to basically the SM value. 

\begin{figure}[t]
\centering
\subfloat[]{%
\label{fig:-a}%
\includegraphics[width=0.45\textwidth]{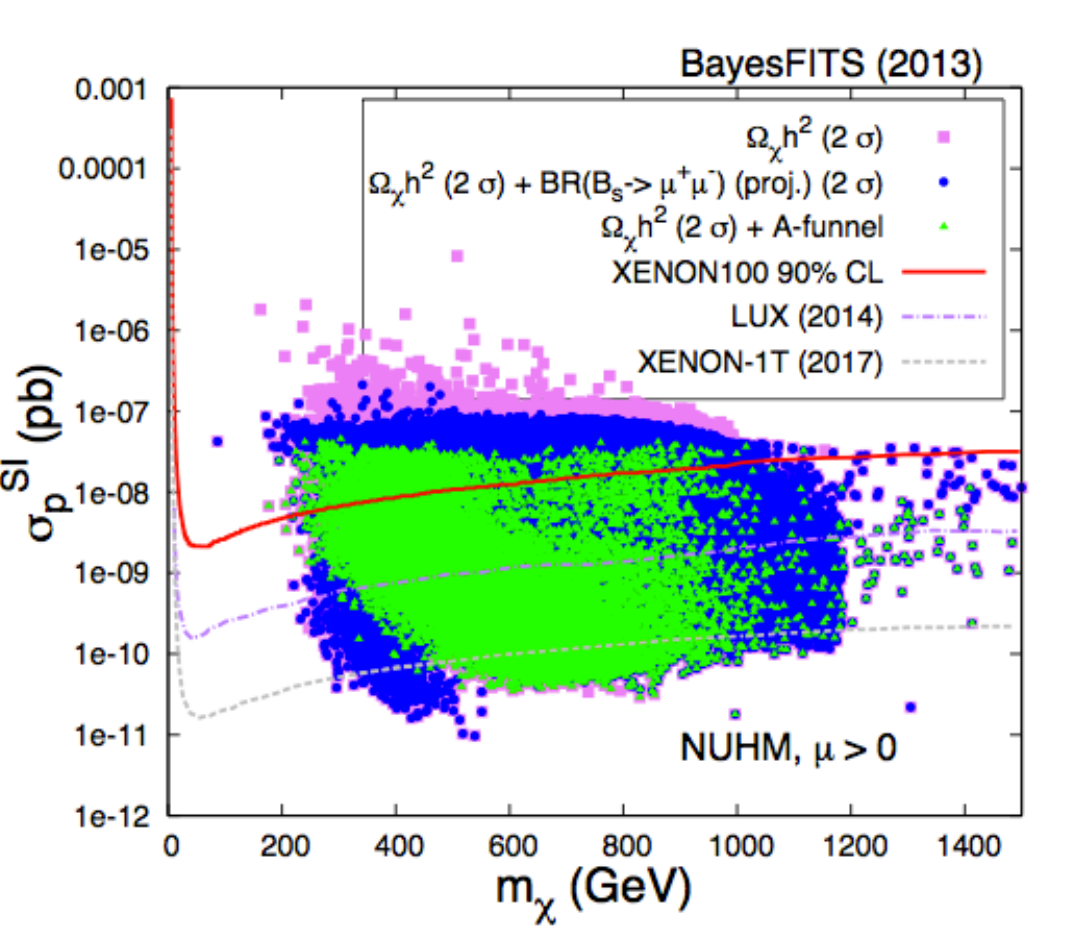}
}%
\subfloat[]{%
\label{fig:-b}%
\includegraphics[width=0.45\textwidth]{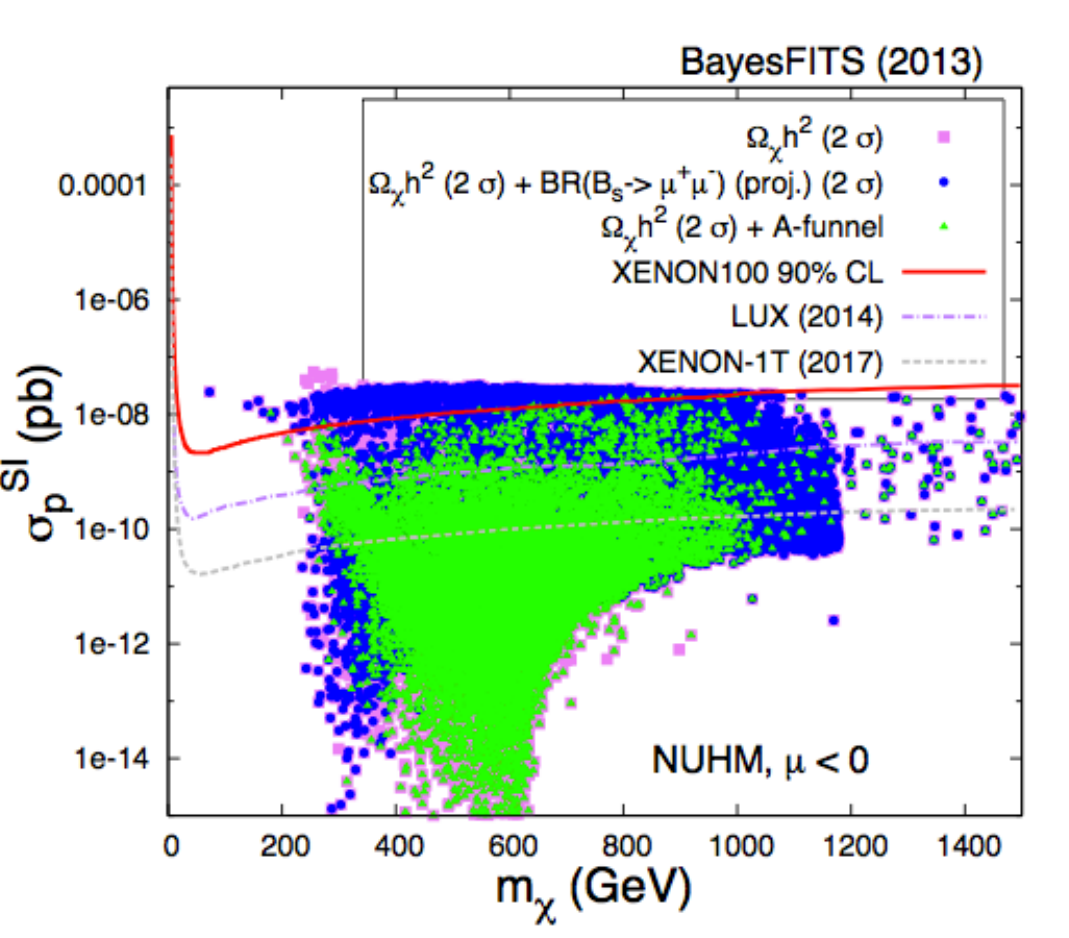}
}%
\caption[]{Scatter plot of the points in the (\mchi, \sigsip) plane of
  the NUHM for \subref{fig:-a} $\mu>0$, and \subref{fig:-b} $\mu<0$,
  that satisfy $\abundchi$ at 2$\sigma$ (pink squares),
  $\abundchi+\brbsmumu_{\textrm{proj}}$ at 2$\sigma$ (blue circles), and $\abundchi$
  at 2$\sigma$ and $|\ma-2\mchi|<100\gev$ (green triangles). The solid red line shows the 90\%~\cl\
  exclusion bound by XENON100 (not included in the likelihood), the dash-dotted purple line the projected sensitivity
  for LUX, and
  the dashed gray line the projected sensitivity for XENON-1T.}
\label{fig:nuhm_bsmu_afun3}
\end{figure} 

The same, unfortunately, is true when it comes to future direct
detection of DM.  In Figs.~\ref{fig:nuhm_bsmu_afun3}\subref{fig:-a}
and \ref{fig:nuhm_bsmu_afun3}\subref{fig:-b} we show the same points
in the (\mchi, \sigsip) plane for $\mu>0$ and $\mu<0$, respectively.
One can see that the green points featuring the AF region with good
relic density and SM-like \brbsmumu\ cover very wide ranges of both
\mchi\ and \sigsip, extending from the SC region to the 1TH region,
which is also present in the NUHM\cite{Roszkowski:2009sm}; for a
recent update see\cite{Strege:2012bt}. One needs to remember that the
limited prior ranges used here for the NUHM (see
Table~\ref{tab:priorsNUHM}) do not fully reproduce the large 1TH
region at $\mchi\simeq 1\tev$. Nevertheless, we added to the plots the
points of the 1TH region that were obtained with extra scans featuring
linear priors in all mass parameters, and much broader ranges (also
given in Table~\ref{tab:priorsNUHM}). This is
allowed, as long as we do not draw any statistical conclusion from the
combination of these chains.  One can see from
Fig.~\ref{fig:nuhm_bsmu_afun3} that it will be much more challenging
to discriminate among the three high probability regions: a DM signal
detected at smaller \mchi\ could be indicative of either the SC or the
AF region while the same at \mchi\ close to 1\tev\ could instead imply
either the AF or the 1TH region. Furthermore, for negative $\mu$, in a
large number of cases with good dark matter relic density and SM-like
\brbsmumu\ (and correct Higgs mass, \etc) accidental cancellations produce \sigsip\ well below
the reach of even one-tonne detectors.

On the positive side, there is one class of DM signal measurements
that could  potentially allow one  to basically rule out the
CMSSM over a whole reasonable range of parameters. The NUHM prominently features a wide region of
roughly $500\gev\lesssim \mchi\lesssim 800\gev$ and \sigsip\
often within the reach of one-tonne detectors which is absent in the CMSSM
(except for a handful of cases with relatively poor \chisq; compare
Fig.~\ref{fig:cmssm_mchi_sig}\subref{fig:-c}). A detection of a signal
in future DM searches indicative of this mass range would then provide
a strong argument against the CMSSM.  


\section{\label{Summary} Summary and Conclusions}

In this paper we have examined the implications from the current and
the projected but realistic sensitivities of both \brbsmumu\ at the
LHC and \sigsip\ in direct DM searches on the CMSSM and the NUHM.
Within the CMSSM we performed an updated global Bayesian analysis of
the CMSSM, with particular focus on the impact of the recent
measurement of \brbsmumu\ at LHCb.  We further extended the parameter
ranges with respect to our previous analysis of the model, and we
updated the limits from CMS direct SUSY searches through our
likelihood map procedure, obtained by simulating the SUSY signal and
the detector efficiencies.  We showed that the same lower bounds apply
to the NUHM as well.  We confirmed that, in the CMSSM, in addition to the
previously identified high posterior probability regions of the
(\mzero, \mhalf) plane favored by the global constraints, at
previously unexplored large CMSSM mass scales a prominent 68\%
credibility region appears, where the LSP is a nearly pure higgsino
with mass of about 1\tev.

We highlighted a correlation between \brbsmumu\ and the $A$-funnel
region of the CMSSM as the above branching ratio and the annihilation
cross section in the AF region both primarily depend on the same
parameters: \mha\ and \tanb.  In this regard, we showed that the AF
region of the CMSSM is at present slightly disfavored (95\%
credibility of the posterior pdf) by the first \brbsmumu\ measurement,
although far from excluded, given the large experimental (and
theoretical) uncertainties. However, with expected future,
significantly reduced uncertainties (experimental of 5\% of the
measured value; theoretical of 5\% of the SM value), this observable
alone will have the potential to basically rule out the whole AF region, and
thus a very broad range of the (\mzero, \mhalf) plane that will for the
most part remain beyond the reach of direct sparticle searches at
the LHC.  Next we showed that DM direct detection search sensitivities
expected for future one-tonne detectors provide a complementary and
strong tool to test and discriminate between the remaining two high
probability regions of the CMSSM: the SC region corresponding to the
LSP mass of $\lsim450\gev$ (and borderline \sigsip) and the $\sim1\tev$
higgsino region with a much wider range of \sigsip. Note also that for $\mu<0$
only the latter case is, for the most part, detectable. This also implies
that a DM signal indicative of the SC region would strongly favor the
positive sign of $\mu$.

The NUHM presents, unfortunately, a much less clear cut behavior with
respect to the interplay of the above observables. In particular this
is so because the pseudoscalar mass can be treated as a free
parameter of the model and can be adjusted, along with the other
parameters, in different ways to yield a good fit to almost all
observables. While high probability regions analogous to the CMSSM are
also present in the NUHM, and no additional ones, they correspond to
different ranges of the parameter space. As a result, unlike in the
CMSSM, one can easily identify the AF region with very SM-like
\brbsmumu. Furthermore, \mchi\ and \sigsip\ in the AF region extend
to much wider ranges than in the CMSSM. For this reason, in the NUHM it
is unlikely to be possible to use future determinations of \brbsmumu\
and \sigsip\ to convincingly rule out the $A$-funnel, which will also
remain for the most part beyond the reach of LHC direct SUSY
searches. On the other hand, a measurement of a DM signal in the mass
range $500\gev\lesssim \mchi\lesssim 800\gev$ would be a strong
indication against the CMSSM where such cases giving a good fit too
all data are absent.

\begin{center}
\textbf{Acknowledgments}
\end{center}
We would like to thank Yue-Lin Sming Tsai for helpful discussions
throughout.  We would also like to thank B.~Allanach for useful
explanations on the details of different versions of SoftSUSY.
L.R. would like to thank G.~Isidori, N.~Mahmoudi, M.~Palutan and
B.~Pietrzyk for correspondence regarding \brbsmumu.  This work has
been funded in part by the Welcome Programme of the Foundation for
Polish Science.  K.K. is supported by the EU and MSHE grant N
POIG.02.03.00-00-013/09.  L.R. is also supported in part by the Polish
National Science Centre grant N N202 167440, an STFC consortium grant
of Lancaster, Manchester and Sheffield Universities and by the EC 6th
Framework Programme MRTN-CT-2006-035505. The use of the CIS computer
cluster at NCBJ is gratefully acknowledged. L.R. is grateful to the CERN
Theory Division for hospitality extended to him during the final stages of this work.

\bibliographystyle{JHEP}

\bibliography{myref}

\end{document}